\documentclass{article}

\usepackage{arxiv}

\usepackage[utf8]{inputenc} 
\usepackage[T1]{fontenc}    
\usepackage{hyperref}       
\usepackage{url}            
\usepackage{booktabs}       
\usepackage{amsfonts}       
\usepackage{nicefrac}       
\usepackage{microtype}      
\usepackage{lipsum}
\usepackage{graphicx,verbatim,float,subfigure,graphicx,color,siunitx}
\usepackage{amsmath}
\graphicspath{ {./images/} }

\usepackage{amssymb}
\usepackage{svg}
\usepackage{multirow}  
\usepackage[export]{adjustbox}
\usepackage{soul}
\usepackage{textcomp}
\usepackage{multirow}
\usepackage{enumitem}

\title{Difference-Based Deep Learning Framework for Stress Predictions in Heterogeneous Media}

\author{
 Haotian Feng and  Pavana Prabhakar$^*$ \\
  Dept. of Civil \& Environmental Engineering \\
  University of Wisconsin-Madison \\
  Madison, WI 53706  \vspace{0.05in} \\
  \texttt{$^*$pavana.prabhakar@wisc.edu}
}

\begin{document}
\maketitle
 
\newcommand{\SPS}[1]{\textcolor{red}{\bf{Sabari: #1}}}

\begin{abstract}
Stress analysis of heterogeneous media, like composite materials, using Finite Element Analysis (FEA) has become commonplace in design and analysis. However, determining stress distributions in heterogeneous media using FEA can be computationally expensive in situations like optimization and multi-scaling. To address this, we utilize Deep Learning for developing a set of novel Difference-based Neural Network (DiNN) frameworks based on engineering and statistics knowledge to determine stress distribution in heterogeneous media, for the first time, with special focus on discontinuous domains that manifest high stress concentrations. The novelty of our approach is that instead of directly using several FEA model geometries and stresses as inputs for training a Neural Network, as typically done previously, we focus on highlighting the differences in stress distribution between different input samples for improving the accuracy of prediction in heterogeneous media. Our DiNN framework consists of three main modules: 1) a sample processing module that calculates the difference geometry and stress contours of each sample with respect to a reference contour extracted from the training samples, 2) an Encoder-Decoder module that predicts stress difference contours using geometry difference contours as input, and 3) a stress prediction module that combines the stress difference contours with the reference contour to construct the final prediction of stress contours. We evaluate the performance of DiNN frameworks by considering different types of geometric models that are commonly used in the analysis of composite materials, including volume fraction and spatial randomness. Results show that the DiNN structures significantly enhance the accuracy of stress prediction compared to existing structures, especially for composite models with random volume fraction when localized high stress concentrations are present.  
\end{abstract}


\section{Introduction}\label{intro}
Stress analysis is an important discipline within engineering, where the primary objective is to determine stresses and strains in structures and materials subjected to external loads. 
Within stress analysis, we typically start with a geometrical description of a structure or material and the expected load acting on it. Typical output of stress analysis is the quantitative distribution of stresses, strains and deformations. There are several approaches for stress analysis of solids, like classical mathematical closed form solutions for partial differential equations, computational simulation, experimental testing or a combination of these methods. Among these, Finite Element Analysis (FEA)\cite{The-FEM,Nonlinear-FEM} is a commonly used computational tool for stress analysis and for designing structures and materials. By reformulating governing partial differential equations (PDE) from strong form to weak form, and implementing these in discrete form within FEA, the response of solids subjected to external loads and boundary conditions can be determined. 

FEA is used extensively for analyzing composite materials\cite{fem-composite-book}, which are heterogeneous and usually made of individual constituent materials with unique properties that are combined together to result in improved physical properties as compared to the individual materials. Composite materials typically consist of matrix and reinforcing constituent materials, where the reinforcing material is stiff and strong, and matrix is made of homogeneous and monolithic material that binds the reinforcements together. Typically, several length scales exist within composites, with reinforcements at the micrometer scale and composites at the meter scale. Detailed FEA at multiple length scales are widely used to analyze and design these composites\cite{EFENDIEV2013116,HOU1997169,FEA-composite}. Although commonly used for stress analysis, FEA can be expensive when used for optimization and multi-scale analysis of structures and materials, especially within composite materials. Hence, the past few years have witnessed a few attempts for substituting traditional FEA with Neural Network (NN), a method within Machine Learning (ML) framework, for structural optimization\cite{3D_autoencoder,Data_driven_topology} and multi-scale analysis\cite{VAMPnets,Self_Learn_molecular}. 

Past researchers have shown great potential for ML and Deep Learning (DL) methods as surrogates for predicting mechanical properties in Computational Solid Mechanics and Computational Fluid Dynamics, without performing Finite Element Analysis (FEA). Earlier attempts to integrate ML techniques into FEA focus on updating model\cite{11-Nourbak}\cite{12-Khadilkar}, defining material constitutive relationship\cite{14-KHOZANI2017441} and approximating nonlinear constitutive behavior\cite{16-DBLP:journals/corr/IsolaZZE16}. With limitations of ML techniques in modelling complex nonlinear models, researchers have resorted to DL method as a surrogate to FEA in the field of stress prediction. Liang et al.\cite{10-LiangDL} introduced image-to-image DL model to substitute FEA for stress distribution estimation in human tissues and proved the feasibility of estimating the linkages between shape features and FEA-predicted results. Guo et al.\cite{10.1145/2939672.2939738} proposed Encoder-Decoder\cite{Encoder-Decoder} structure in CFD to predict non-uniform steady laminar flow in vehicle aerodynamic analysis and showed that it is considerably faster than traditional LBM solvers. Khadilkar et al.\cite{21-ATALLA1998135} proposed two-stream deep learning framework for stress field prediction in 3D printing, by combining extracted 2D features with 3D point clouds using CNN and PointNet\cite{22-inproceedings}. Yang et al.\cite{24-article} proposed an artificial neural network (ANN) to approximate yield function under given boundary conditions. Bhatnagar et al.\cite{Afshar-CFD} improved Encoder-Decoder structure's capability for predicting velocity and pressure field in unseen flow conditions and geometries. More in line with our work, Nie et al.\cite{Nie-Stress-Net} introduced StressNet by implementing Residual Network (ResNet)\cite{He-ResNet} into Encoder-Decoder structure and validated their framework using 2D linear elastic cantilevered structural analysis. This significantly improved the accuracy of von Mises stress field prediction in their models.

Inspired by the ability of CNN to extract high-level features\cite{article} and several previous successful Encoder-Decoder models\cite{seq2seq}, we introduce a new {\bf Difference-based Neural Network (DiNN)} structure in this paper that focuses on geometric and associated stress differences between different samples for stress prediction in composite (heterogeneous) materials. Instead of directly using several Finite Element model geometries and stresses as inputs, DiNN focuses on highlighting the differences in stress distribution between different input samples by using the geometry and stress differences between training samples and a reference model (determined based on training samples) as input data for training the NN. Highlighting these differences in the input sample data is expected to improve the prediction accuracy. This is the first attempt, to our best knowledge, towards predicting stress distributions in heterogeneous media like composite materials that possess severe stress concentrations.

\section{Overview of the Proposed Machine Learning Framework} \label{sec:overview}
\begin{figure}[h!]
	\centering
	\includegraphics[width=\textwidth]{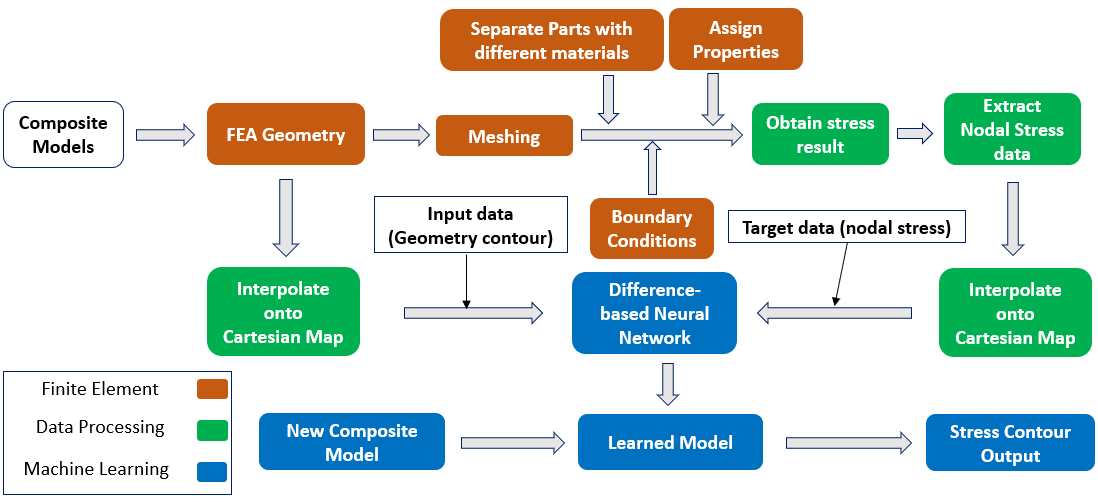}
	\caption{An overview of the proposed ML framework: Composite models are first generated and solved using Finite Element solver to obtain stress distribution contours. The meshed geometry and stress contours are interpolated onto Cartesian Maps and further used for training the DiNN structure. Finally, the trained DiNN is used for predicting stress distribution contours for new composite models. }
	\label{img:MethodOverview}
\end{figure}

In Figure~\ref{img:MethodOverview}, we present an overview of the proposed machine learning framework for composite structures and materials. Throughout the paper, we have used Intel Core™ i7-9700 Processor for performing FEA and post-processing. We perform the machine learning step on NVIDIA GeForce RTX 2080 SUPER with 3072 CUDA cores. We have provided access to our implemented Machine Learning code on our GitHub page and training data on Google Drive, as mentioned in the ``Data Availability'' section at the end of this paper. The GitHub page provides DiNN frameworks for different micromechanical models as well as an example geometry with instructions for running the code to predict stress distribution.

The entire ML framework can be split into three main steps as shown in \mbox{ Figure~\ref{img:MethodOverview}}: 1) data generation using FEA (orange blocks), 2) processing input data for training DiNN (green blocks), and 3) training and testing of DiNN (blue blocks).

\begin{enumerate}
    \item Data generation using FEA
    
First, several composite geometries are randomly generated and meshed using a MATLAB code. They are next numerically solved under quasi-static loading within linear elasticity using the FEA software ABAQUS \mbox{\cite{ABAQUS}} to obtain the spatial stress distribution for each geometry. These geometries and their stress contours will be further used for training the Neural Network in the subsequent steps. Details of the Finite Element Analysis process is described in \mbox{Section~\ref{se:fea_train}}. Mesh convergence study is performed for selecting suitable mesh sizes for different types of geometries considered in this study, such that the stress contour outputs are converged. A brief overview of the FEA is provided in Appendix \mbox{\ref{FEM}}.

\item Processing input data for training DiNN

Next, the meshed geometries and their nodal stress distribution contours that are output from FEA are interpolated onto a uniform global map called Cartesian Map (CM) \mbox{ \cite{Afshar-CFD}} shown in Supplementary \mbox{Figure~\ref{img:CM}}. This enforces the interpolated geometry and the stress contours to be of the same size in order to facilitate NN training and parameter learning through the training steps. In this paper, we use Barycentric coordinate interpolation which provides very stable interpolation compared to other methods like linear interpolation especially in triangular sub-domains. A detailed account of Barycentric coordinate interpolation approach is provided in \mbox{Section~\ref{se:data_processing}}.

\item Training and testing of DiNN

The Difference-based Neural Network (DiNN) is constructed based on an Encoder-Decoder structure, which is discussed in detail in \mbox{Section~\ref{se:ML}}. DiNN includes three modules: sample processing module, Encoder-Decoder module and stress prediction module. Interpolated geometry and stress contours from the Cartesian Map are brought into DiNN for training purposes. During the training process, parameters within DiNN structure are updated using the input training samples. Upon completing the training process, the learned Neural Network further takes unseen/new micromechanical models as input and then outputs the corresponding stress distribution contour of the model as part of testing. 

\end{enumerate}

\section{Model Definition} \label{se:modelDefinition}
\vspace{-0.1in}
\subsection{Introduction to Different Geometric Models Considered}
\begin{figure}[h!]
\centering
\subfigure[]{
  \includegraphics[width=0.2\textwidth]{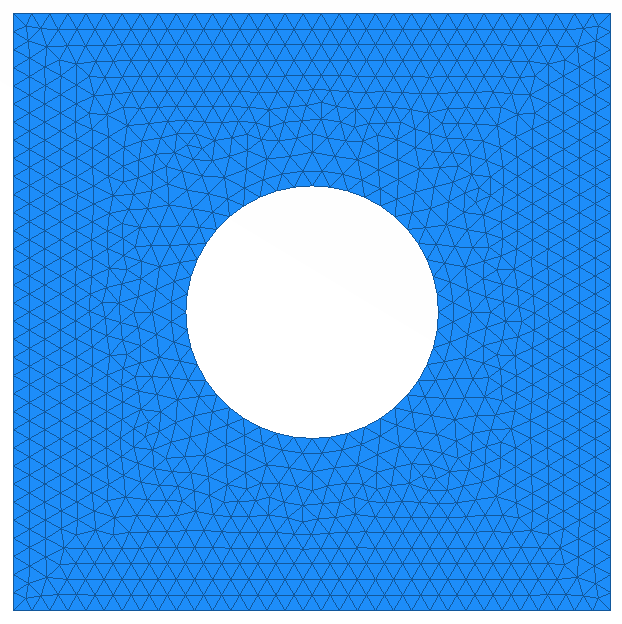}
}
\centering
\subfigure[]{
  \includegraphics[width=0.2\textwidth]{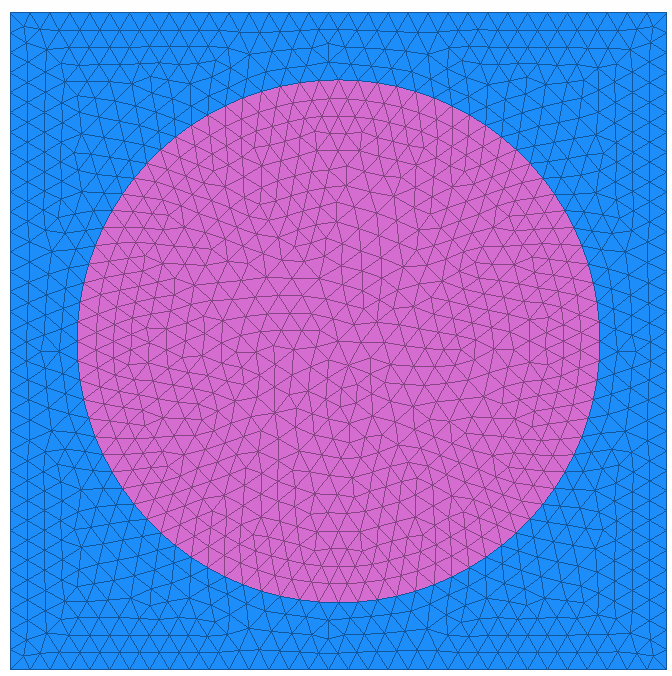}
}
\centering
\subfigure[]{
  \includegraphics[width=0.2\textwidth]{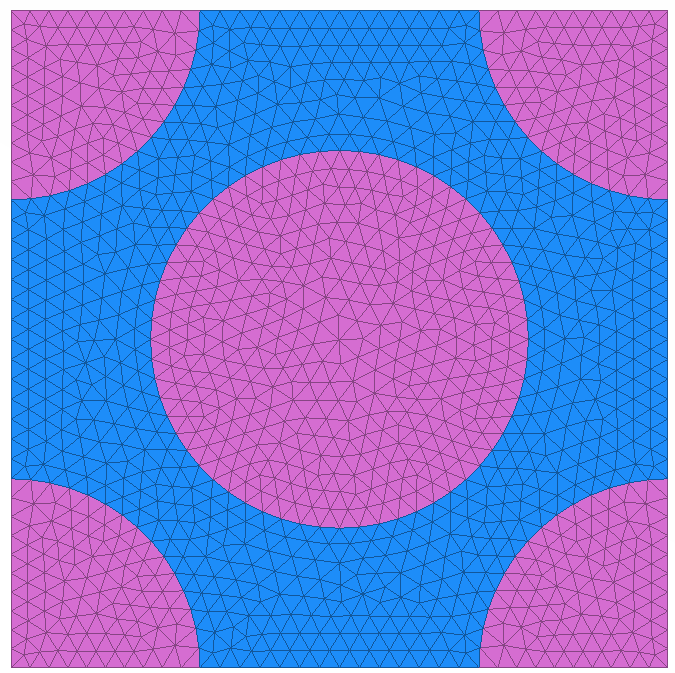}
}
\centering
\subfigure[]{
  \includegraphics[width=0.2\textwidth]{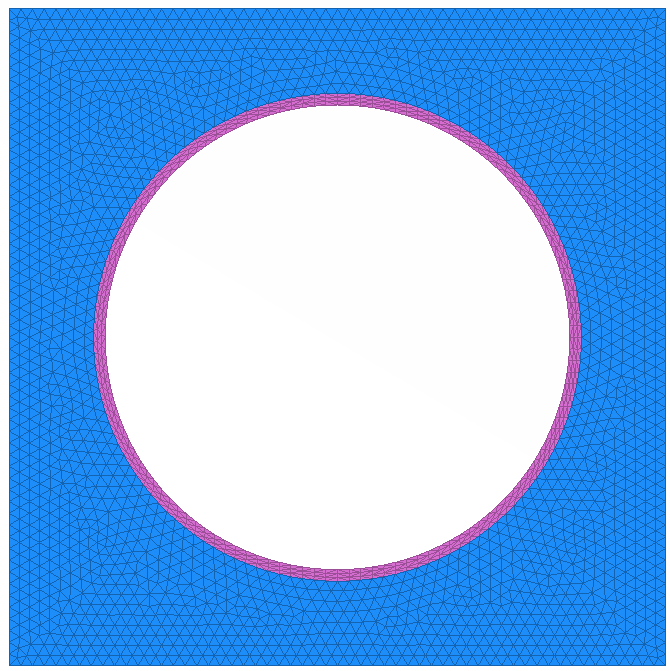}
}
\centering
\subfigure[]{
  \includegraphics[width=0.2\textwidth]{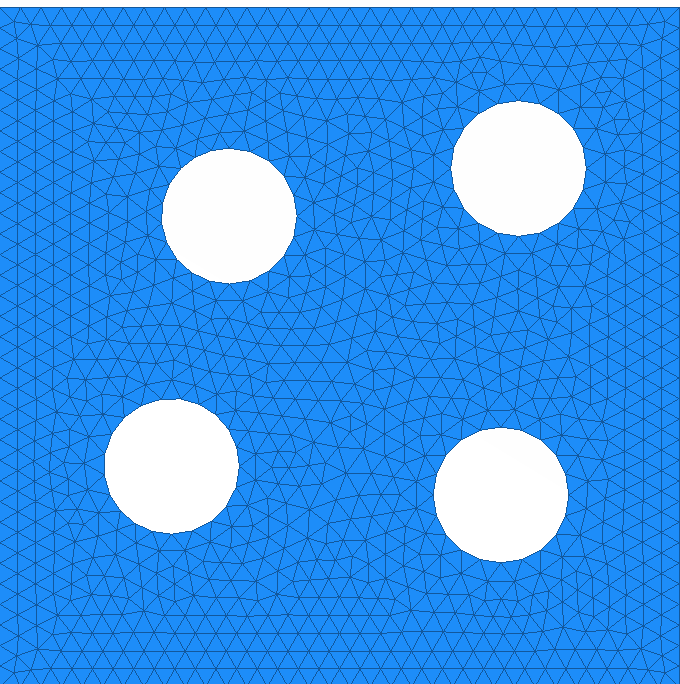}
}
\centering
\subfigure[]{
  \includegraphics[width=0.2\textwidth]{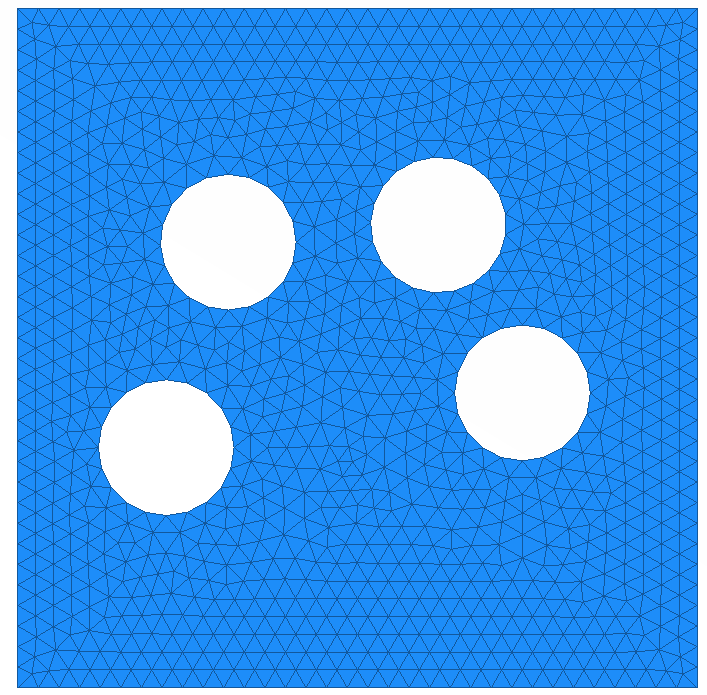}
}
\centering
\subfigure[]{
  \includegraphics[width=0.205\textwidth]{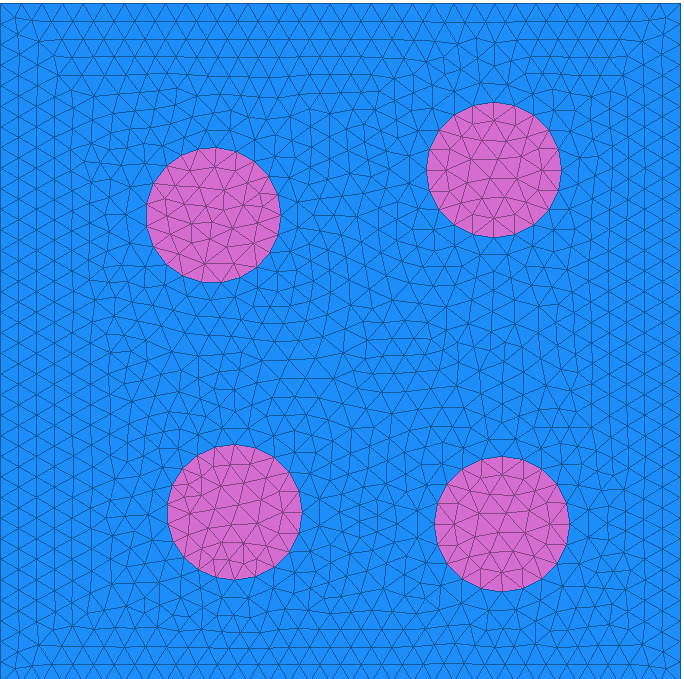}
}
\centering
\subfigure[]{
  \includegraphics[width=0.2\textwidth]{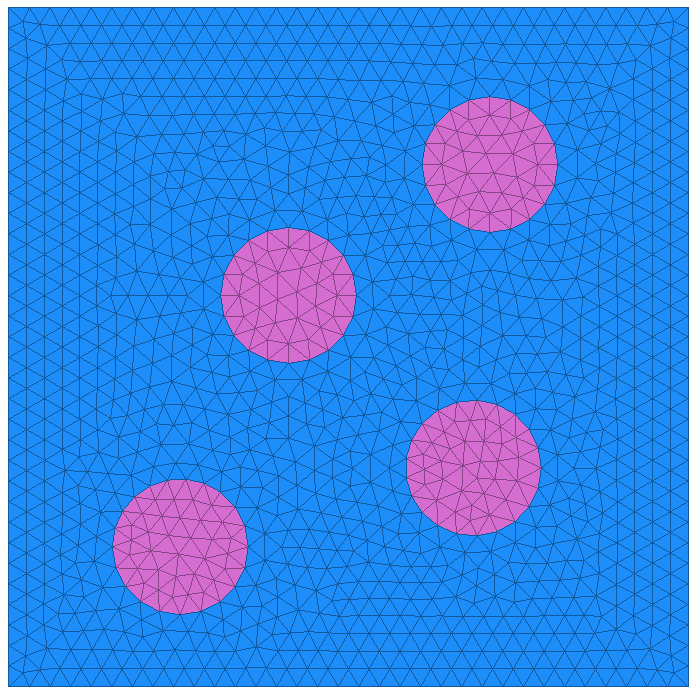}
}
\caption{Meshed composite geometry with volume fraction randomness (top row) and meshed composite geometry with spatial randomness (bottom row): (a) plate with a circular cutout model [PC-VR] (b) square packed fiber reinforced model [SP-VR] (c) hexagonal packed fiber reinforced model [HP-VR] (d) hollow particle reinforced model [HPR-VR] (e) plate with controlled random circular cutout model [PC-SR-C] (f) plate with uncontrolled random circular cutout model [PC-SR-UC] (g) square with controlled random packed fiber reinforced model [SP-SR-C] (h) square with uncontrolled random packed fiber reinforced model [SP-SR-UC]. Blue and pink regions in the geometry models are different materials and the white regions are hollow.}
\label{img:micromech_geometry}
\end{figure}

We consider composite micromechanical models in two categories: volume fraction randomness (VR) and spatial randomness (SR). To account for these variables, we analyze eight different structures shown in Figure~\ref{img:micromech_geometry} as the geometric models in this study: 1) plate with a circular cutout model(PC-VR), 2) square packed fiber reinforced model(SP-VR), 3) hexagonal packed fiber reinforced model(HP-VR), 4) hollow particle reinforced model(HPR-VR), 5) plate with controlled random circular cutout model(PC-SR-C), 6) plate with uncontrolled random circular cutout model(PC-SR-UC), 7) square with controlled random packed fiber reinforced model(SP-SR-C), 8) square with uncontrolled random packed fiber reinforced model(SP-SR-UC). VR models have fixed cutout or fiber centers and random radius, while SR models have random cutout or fiber centers but fixed radius. Overall external dimensions of each model is \SI{10}x\SI{10}{\micro\metre}.

It should be noted that the micromechanical models shown in \mbox{Figure~\ref{img:micromech_geometry}} are computer generated, and are not constructed from real microstructures obtained from microscopy analysis. It is likely that fabrication and specimen preparation performed for microscopy analysis can create defects, debonding, pull out, etc. However, our current models do not consider these additional effects and are used mainly for establishing the approach first. Having said that, our DiNN framework utilizes geometry and stress outputs from FEA models for training the NN. These FEA models are interpolated onto a Cartesian Map forming an image-like matrix, which serve as inputs to the DiNN framework. Hence, as long as we can generate models of real microstructures and perform FEA analysis on them, DiNN is expected to show similar performance. However, the noise introduced from defects may impact the feature extraction step, which may necessitate a higher number of training samples.

\vspace{-0.1in}
\subsection{Importance and Relevance of Each Model Considered}

We consider two dimensional plane stress analysis in this paper and focus on predicting von Mises stresses as both matrix and reinforcement regions have in-plane isotropic properties. In VR models, we vary the geometries by choosing different fiber or cutout radius $r$, which is calculated based on different volume fractions $V_{f}$ using the equation $r=\sqrt{\frac{A_{square}*V_{f}}{\pi}}$. $V_{f}$ describes the volume percentage of one part to the whole model. We randomly generate different values of $V_{f}$ within a target range, where $A_{square}$ is the area of the bounding square region. In SR models, we choose four cutouts or fibers in each model for investigation and validation purposes. Controlled randomness implies that each circle is randomly assigned within each quarter with minimum spacing between circles to be held at 1/2 of circle radius, while uncontrolled randomness implies that only the minimum spacing constraint of 1/2 of circle radius exists. 

\subsubsection{Plate with circular cutout models}

Plate with cutout designs are widely used in mechanical industries, for example airplane cabin window and screw-bolt designs. Under externally applied loads, plate with a cutout experiences high stress concentrations in the vicinity of the cutout. These high stress regions are candidate for localized damage and failure under external loads, and largely affect the mechanical performance of structures. 
A range between 5\% to 25\% for volume fraction is considered for circular cutout models due to relatively small cutout size. Sample meshed models are shown in Figure~\ref{img:micromech_geometry}(a).

In addition to volume fraction, the spatial arrangement of cutout regions in the micromechanical models has a significant factor on the mechanical properties. To investigate Neural Network's performance on models with different spatial arrangements, we consider four random cutouts with controlled and uncontrolled randomness, as shown in Figure~\ref{img:micromech_geometry}(e,f). 

\vspace{-0.1in}
\subsubsection{Fiber reinforced polymer composite micromechanical models}

Fiber reinforced polymer composite (FRPC) materials are widely used in aerospace, automotive, marine and construction industries due to its higher strength comparing to pure polymer matrix. FRPCs typically consist of two parts: stiff reinforcing fibers and a less stiff binding matrix. In this paper, we choose FRPCs as a model system. One key design feature in FRPCs is the fiber volume fraction $V_{f}$, which contributes directly to their mechanical properties. Fibers within FRPCs typically have diameters in the range of few micro meters, for example, carbon fibers are approximately 6 $\mu$m in diameter. Since a structure made of composite is in the order of few meters, representing each fiber in a computational domain is not practical. Often, we resolve to micromechanics for analyzing such composites.

In micromechanical analysis of composites, a large composite domain can be represented by arrays of small repeating unit cells (RUC). Here, we consider two RUCs: square packed and hexagonal packed models of fibers embedded in matrix with fibers having their actual diameter and the entire fiber-matrix domain has a fiber volume fraction equal to that of the macro scale composite domain. Thus, the micromechanics models are in the order of micrometers while effectively capturing the mechanical behavior of large composites. Square-packed RUC of FRPCs has fiber in the center and a square-shaped matrix surrounding it as shown in Figure~\ref{img:micromech_geometry}(b), while hexagonal packed RUC has a full fiber in the center and four quarter fibers in each corner of the square matrix domain as shown in Figure~\ref{img:micromech_geometry}(c). For validation purposes, we simply assume hexagonal packed composite to be a square shaped domain as in the case of a square fiber packed model. Fiber volume fraction of 40\% to 60\% are most common in real FRPC materials. Hence for this paper, we consider the same volume fraction range to generate RUC models with random fiber diameters.

Similar to plate with random cutout model, to account for the effects of spatial arrangement, we further consider square packed composites with randomly distributed fibers as shown in Figure~\ref{img:micromech_geometry}(g,h). 

\vspace{-0.1in}
\subsubsection{Hollow Particle Reinforced Composite}

Hollow particle reinforced composites, also referred to as syntactic foams, are gaining traction in lightweight applications due to their low density, high compressive energy absorption capability and large strains to failure\cite{Breunig2020,Shahapurkar2018,GARCIA2018,Dodda_Microballoon}. Syntactic foams typically consist of stiff hollow particles (often at the micro-scale) randomly dispersed in a softer matrix region, which results in lightweight closed cell foams. A unit cell representation of hollow particle reinforced composite consists of a square block with circular cutout and a thin reinforced ring layer at inner surface of the cutout that represents the cut section of a reinforcing particle. The reinforced layer is typically made of stiffer materials, like glass or ceramics, as shown in Figure~\ref{img:micromech_geometry}(d).
We choose the wall thickness of the reinforcing ring to be approximately 1/22.5 of the inner diameter (hole diameter) based on prior experimental research performed by Jayavardhan et al.\cite{Dodda_Microballoon}. In this paper, we consider the volume fraction of the cutout region to be in the range from 40\% to 60\% with a ring thickness of \SI{0.18}{\micro\metre}.

\vspace{-0.1in}
\subsection{Finite Element Analyses for Training Data Generation} \label{se:fea_train}

All the domains mentioned above are 2D plane stress models with the same external dimensions. Linear elastic mechanical properties of polymer matrix ($E=3.2 \: \mathrm{GPa}, \upsilon=0.31$)\cite{Jones1975MechanicsOC} are assigned, which are typical of epoxy resin used in fiber reinforced polymer matrix composites. For reinforced composites, linear elastic carbon fiber properties ($E = 8.0 \: \mathrm{GPa}, \upsilon = 0.35$)\cite{PRABHAKAR2013-1,PRABHAKAR2013-2,PRABHAKAR2013-4,High_E_CF_stanage} in the plane perpendicular to the fiber direction are assigned. The external boundaries of each domain are defined as $\Gamma_1$, $\Gamma_2$, $\Gamma_3$ and $\Gamma_4$, respectively, for the top, left, bottom and right edges. The boundary conditions on each boundary is described in terms of horizontal ($u$) and vertical ($v$) displacements, where $v = -\SI{0.1}{\micro\metre} \; \mathrm{on} \: \Gamma_1$, $u = 0 \; \mathrm{on} \: \Gamma_2$, $v = 0 \; \mathrm{on} \: \Gamma_3$ and $u = \mathrm{constant} \; \mathrm{on} \: \Gamma_4$. Essentially, each model is subjected to a positive displacement along the vertical direction subjecting them to compression.

Using the above mentioned inputs to each domain, we perform mesh convergence analysis to determine the maximum (max) mesh size that provides converged stress predictions. From this analysis, we identify that the hollow particle reinforced composite model requires a max mesh size of \SI{0.2}{\micro\metre} and the other three models require a max mesh size of \SI{0.3}{\micro\metre}. By assigning material properties along with above-mentioned boundary conditions, we generate stress distribution contours using FEA static solver.

\section{Difference-based Neural Network Framework} \label{se:ML}

\subsection{Framework Overview}
\begin{figure}[h!]
	\centering
	\includegraphics[width=0.9\textwidth]{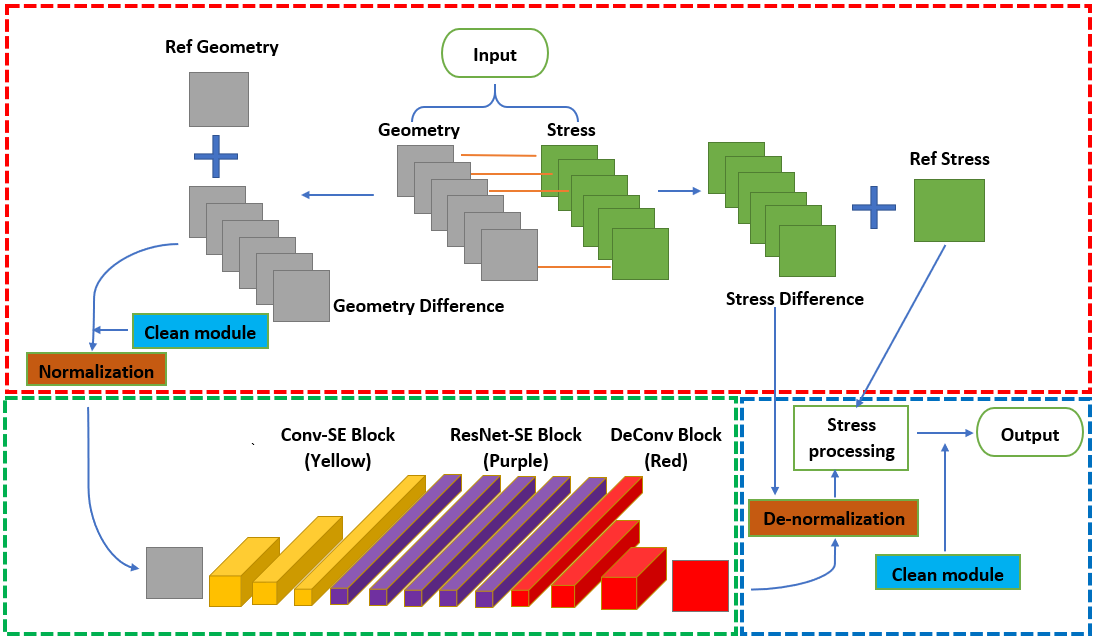}
	\caption{Difference-based Neural Network Framework: sample processing module (red dash line), Encoder-Decoder module (green dash line), stress prediction module (blue dash line). Two orange blocks are added for DiNN-N structure and two blue blocks are added in addition for DiNN-NC structure.}
	\label{img:NeuralNetwork}
\end{figure}

An Encoder-Decoder Neural Network\cite{Encoder-Decoder} is typically trained using geometry and stress contours directly as inputs. Past researchers have demonstrated that embedding residual learning into Encoder-Decoder structure can significantly improve the accuracy of prediction. Nie et al.\cite{Nie-Stress-Net} proposed StressNet structure based on residual learning algorithm, which was shown to improve the accuracy of stress distribution prediction. However, the existing StressNet structure does not provide high prediction accuracy when localized high stresses exist, like stress concentration, especially within composite materials. In this paper, a novel NN structure is developed that embeds engineering and statistical knowledge for stress prediction. During engineering design iterations, engineers typically use an initial design as the reference model and then refine the subsequent designs. Similar to this idea, a known geometry model and corresponding stress distribution contour are chosen as the reference model when training the target NN. Then, the differences between different geometry contours are used to guide NN to focus on training stress difference contours $\sigma$. We refer to this as a Difference-based Neural Network (DiNN) structure shown in Figure~\ref{img:NeuralNetwork}.

The base DiNN structure shown in \mbox{Figure~\ref{img:NeuralNetwork}} mainly consists of three modules: sample processing, Encoder-Decoder and stress prediction. Sample processing module (see Section \mbox{\ref{subsec:processing-of-training-example}}) is the first module in the DiNN structure, which takes the geometry and corresponding stress contours as inputs and performs data preprocessing before training the NN. Encoder-Decoder module (see Section \mbox{\ref{encoder-decoder}}) is the second module in the DiNN structure, which takes geometries in 2D space that are processed after sample processing module and extracts high level features through several Conv-SE blocks. These high level features are then arranged in 3D space, which are further strengthened by ResNet-SE blocks. Finally, these features are collapsed back to original 2D space using Deconv blocks. The last module is the stress prediction module (see Section \mbox{\ref{subsec:final-predict-module}}), which combines the outputs from the Encoder-Decoder module with the reference stress contour extracted in the sample processing module to predict the final stress contours. 

Difference-based Neural Network with normalization (DiNN-N) is built upon DiNN by adding additional normalization and de-normalization blocks (orange color). Further, Difference-based Neural Network with normalization and clean module (DiNN-NC) is developed based on DiNN-N structure by adding two additional clean modules (blue color). An example of how DiNN predicts the stress distribution contours on a square packed composite model is shown in Supplementary Figure~\ref{img:NN-flowchart}.

\subsection{Sample processing module} \label{subsec:processing-of-training-example}
Sample processing module mainly extracts geometry and stress contour information from the training data. Mean geometry and stress contours across training models are extracted as reference sample (labelled as Ref Geometry and Ref Stress). Geometry difference contours, stress difference contours and mean stress contours are used further for training the NN. To avoid covariate shifting and improve training efficiency, a pair of normalization and denormalization modules are added. Denormalization module is defined in Section~\ref{subsec:final-predict-module}. Normalization block is developed based on Min-Max feature scaling function\cite{Min-Max} described in Equation~\ref{eqn:norm} as follows: 
 
{
\vspace{-0.2in}
\begin{equation}
    \mathrm{Normalization: output} = \frac{\mathrm{input}-\min(G)}{\max(G)-\min(G)}
\label{eqn:norm}
\end{equation}
}

\noindent where, $G$ refers to the labelled geometry difference contour.

\subsection{Encoder-Decoder module}
\label{encoder-decoder}
\begin{figure}[h!]
\vspace{-0.1in}
\centering
\subfigure[]{
  \includegraphics[width=1\textwidth]{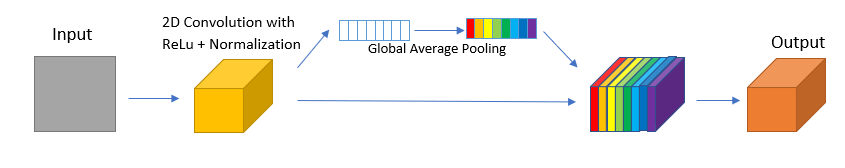}
}
\centering
\subfigure[]{
  \includegraphics[width=1\textwidth]{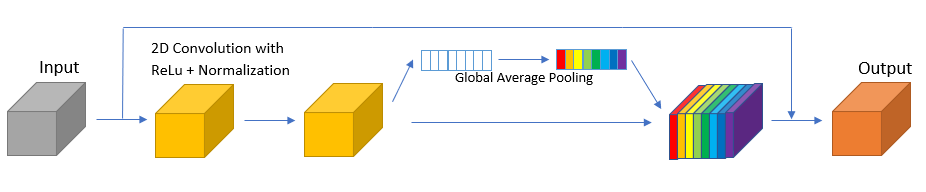}
}
\centering
\subfigure[]{
	\includegraphics[width=1\textwidth]{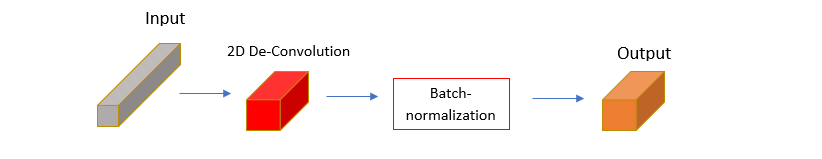}
}
\vspace{-0.1in}
\centering
\subfigure[]{
	\includegraphics[width=0.7\textwidth]{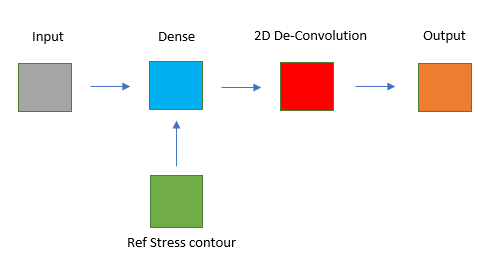}
}
\caption{Inner structures of (a) Conv-SE block; (b) ResNet-SE block; (c) Deconv block; (d) Stress processing block}
\label{img:sub_modules}
\end{figure}

Encoder-Decoder module consists of three types of blocks: Conv-SE (yellow), ResNet-SE (purple) and DeConv (red) blocks: (1) Each Conv-SE block, as shown in Figure~\ref{img:sub_modules}(a), consists of one 2D convolutional layer with ReLU and Batch-normalization, and one Squeeze-and-Excitation (SE) block\cite{hu2018senet}. The convolutional layer extracts high level key features of geometry difference contours and SE block adaptively re-calibrates channel-wise feature responses by modelling inter-dependencies between different channels. (2) ResNet-SE block is constructed based on ResNet architecture and consists of two convolutional blocks and one SE block to enhance the extracted inter-dependent high level features, as shown in Figure~\ref{img:sub_modules}(b). (3) DeConv blocks follow ResNet-SE blocks, and each consist of one 2D deconvolutional layer\cite{NIPS2014_5485} with Batch-normalization, as shown in Figure~\ref{img:sub_modules}(c). The DeConv blocks expand key features and finally back to the original input dimensions. Since the differences between the original and mean contours generally have a zero mean and certain variations, we assume that it follows a Gaussian distribution, and hence use Glorot initialization for weight initialization\cite{Glorot10understandingthe}. 

\subsection{Stress prediction module} \label{subsec:final-predict-module}
Stress prediction module shown in Figure~\ref{img:NeuralNetwork} follows the Encoder-Decoder module. The De-normalization block reverts the effect of normalization, as defined in Equation~\ref{eqn:denorm}. 

\begin{equation}
    \mathrm{Denormalization: output} = \mathrm{input}*(\max(\sigma)-\min(\sigma)) + \min(\sigma)
\label{eqn:denorm}
\end{equation}

\noindent Here, $\sigma$ represents the stress difference contour. A stress processing block is added after the De-normalization block, which generates the predicted stress contours as shown in Figure~\ref{img:sub_modules}(d). It consists of one dense block for combining the predicted stress differences (input) and the reference stress contour, followed by 2D De-Convolution block to smooth the prediction. The kernel size of the De-convolution layer is selected based on the uniformity in stress distributions. A kernel size of [2,2] is used for the De-Convolution block if large stress concentrations occur (DiNN-NC) and a kernel size of [1,1] is used otherwise (DiNN-N). Using [2,2] kernel size is beneficial in models with large stress concentrations as the differences between pixel values are higher. Here, we consider that large stress concentration exists if the stress ratio defined as $R_{\sigma} = \sigma_{max}/\sigma_{mean}$ is larger than 2.
\vspace{-0.2in}

\subsection{Clean module (Only for geometry with cutout region)}
For geometric models that have regions of no material, like in the case of plate with circular cutout model and hollow particle reinforced model, varying the size or position of the cutout regions can introduce undesired negative values during subtraction with the reference contours (refer to Section~\ref{subsec:statistical-property-analysis}). Hence, we introduce an additional module called ``clean module''. This module performs element-wise multiplication between the target contour and material contour $M$, whose regions with material are labelled as '1' and regions without material (cutout) as '0'. The multiplication can be represented with Hadamard product\cite{Hadamard} shown in Equation~\ref{eqn:clean_module}, manually forcing regions without material to be zero valued. 

{
\vspace{-0.2in}
\begin{equation}
    C_{ij}=(A \circ M)_{ij}=A_{ij}M_{ij}
\label{eqn:clean_module}
\end{equation}
    
\[ M_{ij}=
    \begin{cases}
    1, & \text{if material exists at a node}\\
    0, & \text{if material is absent at a node}
    \end{cases}
\]
}

\noindent Here, $A_{ij}$ represents the input contour and $C_{ij}$ represents the output contour after passing through the clean module. Symbol $\circ$ refers to the Hadamard product.


The key difference between DiNN-N and DiNN-NC is the addition of a clean module within DiNN-NC. Specifically, DiNN-N structure can be viewed as a special case of DiNN-NC, where the material contour $M$ in Equation~\ref{eqn:clean_module} is a matrix with all-ones as no cutout region exists. Besides, to account for the presence of a clean module, we choose difference kernel sizes, as discussed in Section~\ref{subsec:final-predict-module}. 

\section{Interpolation on Cartesian Map} \label{se:data_processing}

\subsection{Data Pre-processing: Barycentric Coordinate Interpolation} \label{se:BC_process}
Meshed geometries and corresponding stress contours are used for training and evaluating different NN frameworks. Each training sample has a unique shape due to geometry randomness considered, resulting in different meshes in the domain. This poses difficulty for NN training. To render the geometry as well as stress contour trainable for NN, these contours are further interpolated onto a global Cartesian Map (CM), such that all contours have the same size. CM has the same dimensions as that of the geometric models, which are \SI{10}{\micro\metre}-by-\SI{10}{\micro\metre}. 

Triangulation-based linear interpolation\cite{Tri-interp-Amidror} is widely used for map-to-map interpolation due to its simplicity and efficiency. With known position of each node in the CM, this algorithm searches for three nearest nodes in the original Finite Element mesh to form a triangle. This could introduce large artificial errors due to matrix illness in FEA meshed models when complex boundaries or stress concentrations are present that may require non-uniform mesh with smaller mesh size. To avoid this in the DiNN framework, we use Barycentric Coordinate (BC) system\cite{Barycentric}, also known as area coordinates, which normalizes each axis and generates homogeneous coordinates. Due to this characteristic, BCs are extremely useful in rendering the interpolation more stable within triangular sub-domains. BC can accurately determine nodal location with respect to triangular mesh, as well as interpolation coefficient with three vertices. Each node in the CM is projected onto FEA mesh model to find the triangle it falls within. Then, the target value of each node on the CM is calculated using corresponding values of three vertices of the triangle. Relative positions between a node and a triangle are determined by three Barycentric parameters $\lambda_{1}, \lambda_{2}, \lambda_{3}$, which are determined using Equation~\ref{eqn:bc_param} as:

\begin{subequations} 
\begin{equation}
    \lambda_{1}=\frac{(y_{2}-y_{3})(x-x_{3})+(x_{3}-x_{2})(y-y_{3})}{(y_{2}-y_{3})(x_{1}-x_{3})+(x_{3}-x_{2})(y_{1}-y_{3})} \\
\end{equation}
\vspace{-0.3in}
\begin{equation}
    \lambda_{2}=\frac{(y_{3}-y_{1})(x-x_{3})+(x_{1}-x_{3})(y-y_{3})}{(y_{2}-y_{3})(x_{1}-x_{3})+(x_{3}-x_{2})(y_{1}-y_{3})} \\
\end{equation}
\vspace{-0.3in}
\begin{equation}
    \lambda_{3}=1-\lambda_{1}-\lambda_{2}
\end{equation}
\label{eqn:bc_param}
\end{subequations}

\noindent Here, $\{x_{i},y_{i}\}$ represents the coordinate of each vertex of the triangle and $\{x,y\}$ represents the coordinate of a target node on the CM. Relative position of a node with respect to vertices of the triangle can be visualized in Supplementary Figure~\ref{img:Rela_pos}. 

Interpolated nodal values on the CM can be determined by Equation~\ref{eqn:S_lambda} as:
\begin{equation}
    S=\lambda_{1}S_{1}+\lambda_{2}S_{2}+\lambda_{3}S_{3}
\label{eqn:S_lambda}
\end{equation}
where, $S$ is the nodal value on the CM containing geometry label or physical information like stress. $S_{i}$ is the $i^{th}$ nodal value of the triangular element within which the node on the CM falls.

\begin{figure}[h!]
\centering
\subfigure[]{
  \includegraphics[width=0.32\textwidth]{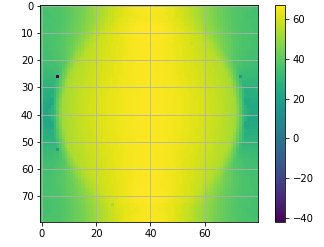}
}
\centering
\subfigure[]{
  \includegraphics[width=0.3\textwidth]{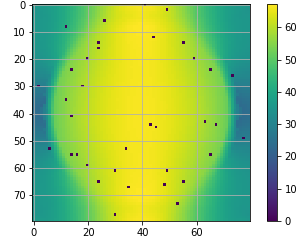}
}
\centering
\subfigure[]{
  \includegraphics[width=0.3\textwidth]{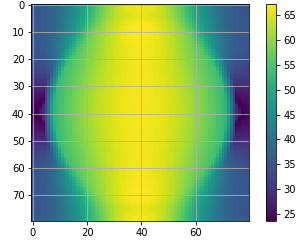}
}
\caption{Interpolation of SP-VR model's stress contour onto Cartesian Map using (a) linear interpolation with nearest three nodes (b) linear interpolation with nearest five nodes (c) Barycentric coordinate interpolation}
\label{img:stress-interpolation}
\end{figure}

Next, we perform a comparison between von-Mises stress contours obtained using linear interpolation (with nearest three nodes and nearest five nodes) and Barycentric coordinate interpolation on a square packed model. It is observed (Figure~\ref{img:stress-interpolation}) that the nearest three nodes linear interpolation generates negative values, while nearest five nodes linear interpolation also manifests several noisy points with minimum stress approximately equal to zero. On the other hand, BC interpolation does not cause illness and singularity, and successfully represents the stress distribution contour and stress ranges after interpolation.

\begin{figure}[h!]
\centering
\subfigure[]{
  \includegraphics[width=0.2\textwidth]{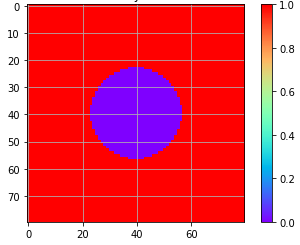}
}
\centering
\subfigure[]{
  \includegraphics[width=0.5\textwidth]{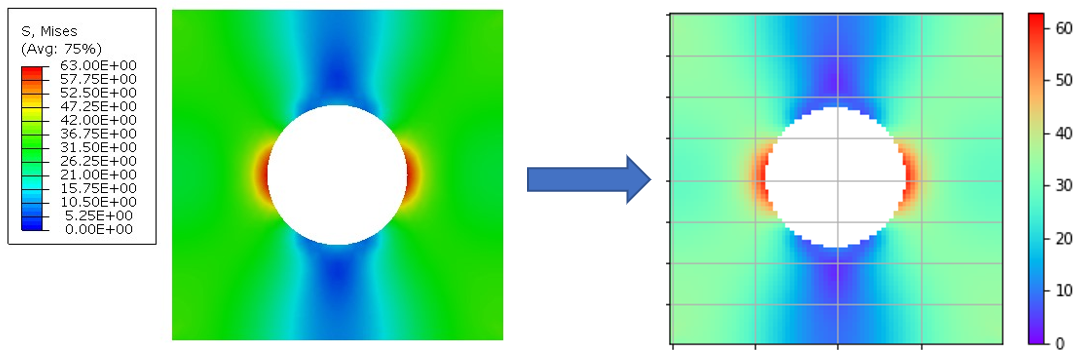}
}
\hspace{0.5in}
\centering
\subfigure[]{
  \includegraphics[width=0.2\textwidth]{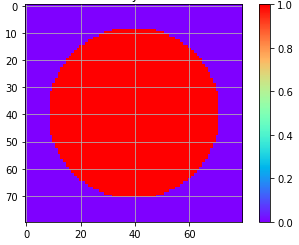}
}
\centering
\subfigure[]{
  \includegraphics[width=0.5\textwidth]{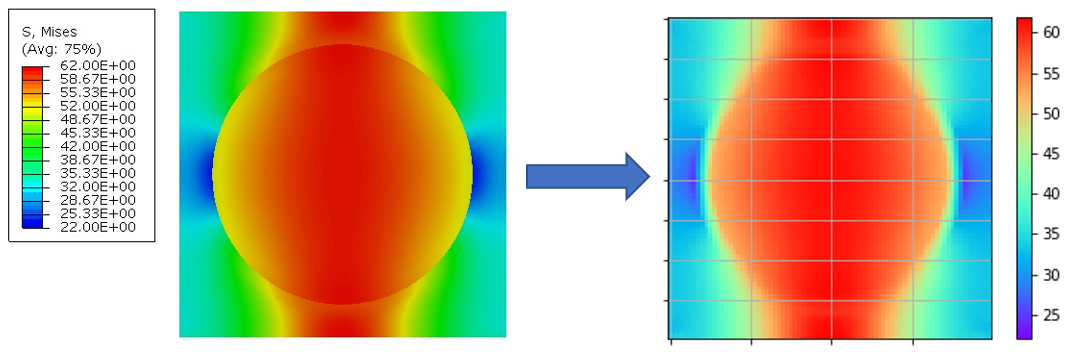}
}
\caption{Geometry contour labelling and Stress interpolation from FEA output to Cartesian Map for PC-VR (a and b) and SP-VR model (c and d).}
\label{img:geometryLabel_stress}
\end{figure}

In addition to the interpolation method, CM density is a key factor that contributes towards interpolation values. To ensure that the CM captures important statistical features of the geometry and stress distribution contours, we perform stress interpolation analysis to establish the CM density that gives reasonable interpolation accuracy and efficiency. From this analysis (Supplementary Figure~\ref{img:CM_interp}), we observe that the interpolation accuracy increases with increasing CM density, while the interpolation speed decreases. To strike a balance between accuracy and efficiency, especially accounting for large sample sizes, we select a CM density of 79-by-79, which has the shape of 80-by-80 after converting to nodal matrix. Such CM density can provide an interpolation accuracy above 99\% in max stress and a reasonable interpolation speed of 26 samples/minute. Examples of geometry and stress contour interpolation onto CM for PC-VR and SP-VR models are shown in Figure~\ref{img:geometryLabel_stress}.

\subsection{Statistical Property Analysis} \label{subsec:statistical-property-analysis}

As compared to the existing StressNet structure, the objective of our proposed DiNN structure is to improve the accuracy of prediction based on a reference data set and render the training process more statistically stable. To that end, we calculate the mean and skewness in the training samples by considering a subset of the data points in the regions where the geometry of the models change (Supplementary Figure~\ref{img:statistic-point}), that is in the vicinity of cutout or fiber. We consider nine points (A-I) in this region and calculate skewness values based on Pearson's second skewness coefficient shown in Equation~\ref{eq:skewness} \cite{math-stat}.
\begin{equation}\label{eq:skewness}
    \mathrm{skewness} = \frac{3(\mathrm{mean}-\mathrm{median})}{\mathrm{standard \; deviation}}
\end{equation}

Skewness in training data can cause imbalance problem and eventually reduce the prediction accuracy of NN by introducing unbalanced data \cite{class_imbalance,DBLP:journals/corr/abs-1305-1707}. There are typically two ways of reducing skewness: 1) model-oriented (reduce skewness in the structure) and data-oriented (reduce skewness by pre-processing)\cite{SUBBANARASIMHA2000117}. To reduce the skewness in data, element-wise root operation on the training samples is performed for positive skewed data and element-wise power is used to deal with negative skewed data\cite{skew-transform}. In this section, we use three models for illustration: PC-VR, SP-VR and PC-SR.

Since geometry labels are constant for individual regions of heterogeneous media, nodal stress values are the main source of skewness. Supplementary Table~\ref{tab:max-min} shows how DiNN structures change max-min ranges for the three models investigated. DiNN structures make the training data more zero centered and symmetric. For models with cutout, DiNN-NC reduces the max-min range that can increase due to the presence of cutout region. DiNN-NC is more effective for models with significant stress concentration near the cutout edges. Supplementary Tables ~\ref{tab:plate-cut-stat},~\ref{tab:square-stat} and \ref{tab:plate-random-stat} summarize the mean, median and skewness values for 1000 different models. For SP-VR, since DiNN only performs mean contour subtraction (a linear operation), the data has a zero mean value and skewness value will not change. For PC-VR and PC-SR-UC models, by adding clean module within DiNN, DiNN-NC alters the statistical properties for cutout regions, resulting in few nodes with non-zero mean values. On the other hand, clean module switches all skewness values to positive as compared to the original sample. This helps in improving the accuracy of DiNN prediction, and facilitate further steps to reduce the effect of skewness if needed.

\section{Machine Learning Model Inputs}\label{se:compModel}

As discussed in the previous section, both nodal stress contours and geometry contours for different models are interpolated onto CM for training the NN. Unique labels are assigned to each region, including cutout region, fiber region, particle ring region and matrix region. In this paper, same boundary conditions and loads are maintained in all analyses, and hence, the effects of boundary condition labelling are not discussed.

To train the NN and test its performance, we randomly generate 2000 samples for each geometry considered based on different cutout or fiber volume fractions. When training for each model, 2000 samples each are randomly split into 80\% for training with 10\% for cross-validation and 10\% for testing. Random split seed is controlled based on the pseudorandom number generator\cite{pseudo-number} developed in scikit-learn\cite{scikit-learn} for comparison purposes.

\subsection{Geometry labelling method}

To identify different regions in each model and facilitate machine learning process, we need to assign different labels to each region. Overall labelling method is summarized in Equation~\ref{eqn:geom_label} as:

\begin{equation}
    G_{ij}=
    \begin{cases}
    0,  & \text{if any material is absent at CM node}\\
    1,  & \text{if matrix material exists at CM node}\\
    2,  & \text{if reinforcement material exists at CM node}
    \end{cases}
\label{eqn:geom_label}
\end{equation}

\noindent where, $G_{ij}$ represents the label for a target geometry.

\subsection{Analyzing the Effect of Skewness}\label{subsec:analyze-skewness}

We previously discussed the effect of skewness in training data on the prediction accuracy of NN in Section~\ref{subsec:statistical-property-analysis}. We noticed that all the skewness values become positive after adding the clean module to DiNN. To determine the optimal solution for reducing the effect to positive skewness and enhancing the prediction accuracy, we introduce a skewness correction factor $p$. By considering different root 'p' over the stress contour values prior to training the NN, our goal is to reduce the influence of skewness on the accuracy of prediction. Element-wise root values are calculated before extracting the mean stress contour and correspondingly element-wise power is introduced after obtaining the prediction from the Deconv block. The predicted stress contour can be expressed using Equation~\ref{eqn:sigma_p}, where the original input geometry contour to the NN is $G$, $NN(w,b)$ is the NN with parameters $w$ and $b$. $\sigma_{average}$ is the average stress contour used as the reference contour during training, which is calculated using Equation~\ref{eqn:sigma_p_ave}. 

\vspace{-0.3in}
\begin{subequations}
\begin{equation}
    \sigma_{output,ij} = [(G * NN(w,b))_{ij} + \sigma_{average,ij}]^{p}  \quad \quad (0<p\leq1) \label{eqn:sigma_p}
\end{equation}
\vspace{-0.3in}
\begin{equation}
    \sigma_{average,ij} = \frac{1}{N}\sum_{n=1}^{N}\sqrt[p]{\sigma_{n,ij}} \label{eqn:sigma_p_ave}
\end{equation}
\end{subequations}

Different values of $p$ are considered for investigating the hidden relationships within DiNN using HP-VR model. The results of this analysis is discussed in Appendix~\ref{sse:skewness}.

\section{Results and Discussion}\label{se:Results}

We construct NN frameworks within Tensorflow 2.0.0 and train it on GPU as discussed in Section~\ref{sec:overview}. To test the prediction capability of different NN frameworks for eight different geometric models considered, we use the StressNet structure for comparison. The accuracy of stress prediction within each component of composite materials, NN training duration and training loss are used for evaluating each model.

\subsection{Prediction error definition}

We evaluate the prediction accuracy of max stress based on the max stress error rate (MER) as defined in Equation~\ref{eqn:MER} and evaluate the training loss based on the mean squared error (MSE) in stress prediction as defined in Equation~\ref{eqn:MSE}.

\vspace{-0.2in}
\begin{subequations}
\begin{equation}
    MER=\frac{1}{N}\sum_{i=1}^{N}\frac{|\mathrm{max}({\hat{Y}_{i}})-\mathrm{max}({Y_{i}})|}{\mathrm{max}({\hat{Y_{i}}})}\times100 \%
\label{eqn:MER}
\end{equation}
\vspace{-0.2in}
\begin{equation}
    MSE=\frac{1}{N}\sum_{i=1}^{N}{(\frac{1}{n}\sum_{j=1}^{n}{(Y_{i,j}-\hat{Y}_{i,j})}^2)}
\label{eqn:MSE}
\end{equation}
\end{subequations}

\noindent Here, $N$ is the total number of samples in the testing set and $n$ is the total number of nodes in the CM. $Y_{i}$ is the nodal stress predicted using the NN and $\hat{Y}_{i}$ is the ground truth nodal stress mapped onto the CM from FEA. $Y_{i,j}$ and $\hat{Y}_{i,j}$ represent the $j^{th}$ nodal stress values in $i^{th}$ sample obtained from the NN and FEA, respectively.

MER estimates the maximum stress prediction error within a model, which is relevant for evaluating composite model's fracture or plasticity initiation. MSE evaluates the accuracy of the stress distribution contour prediction in the entire model, which is important for estimating effective properties, such as the effective stiffness. Training accuracy and efficiency of different NN structures on eight geometric models considered in this paper are summarized in Section~\ref{sse:NNprediction}.

\subsection{Neural Network prediction results} \label{sse:NNprediction}

We evaluate different NN frameworks, including a baseline model (StressNet) and three types of difference-based structures developed by us: DiNN, DiNN-N and DiNN-NC, on all geometric models considered in this paper. DiNN-NC structure was used only when cutout region exists in the target geometry. 

We train each NN framework with 1000 and 2000 samples separately, and use stochastic gradient descent (SGD)\cite{bottou_1999} as the optimizer with learning rate set as 0.001. We choose the training epoch number and steps when prediction accuracy has converged. Supplementary Figures~\ref{img:hollow_loss} and ~\ref{img:square_loss} show examples of training loss profile for PC-VR and SP-VR models with respect to training epochs on a linear and log scale with different sample sizes. These plots indicate that the loss converges within given epochs. Figures~\ref{img:square_duration_predict} to~\ref{img:particle_duration_prediction} show examples of predicted stress contours compared to that generated by FEA for each geometric model as well as the errors (MER and MSE) and training duration while considering 1000 total number of samples. Prediction results for 1000 total samples are used for illustration purposes. Based on the training results obtained from these eight geometric models, we conclude the following:

\begin{enumerate}[label=(\Alph*),noitemsep]
\setlength{\itemsep}{0pt}

\item {\bf Prediction accuracy} for models without cutout regions:

\begin{itemize}[leftmargin=0in,noitemsep]
\setlength{\itemsep}{0pt}
\item Table~\ref{tab:pred_error_no_cutout} shows the prediction performance comparing DiNN-N and baseline model StressNet for 1000 samples. We observe that DiNN-N increases the prediction accuracy in both MER and MSE for SP-VR and HP-VR models. DiNN highlights the changes in geometry and stress distribution between different training models and focuses on training the stress difference contours. This helps the Neural Network improve the prediction accuracy especially in key areas instead of devoting resources towards training the whole stress distribution. This is discussed in detail in \mbox{Section~\ref{se:ML}}. In addition, within the DiNN structure, the Conv-SE blocks have SE blocks following each convolutional layer to improve the quality of extracted high level features compared to StressNet. Besides, the combination of normalization and denormalization modules added in the DiNN-N structure reduces covariate shifting that results in improving the prediction accuracy, as discussed in \mbox{Section~\ref{subsec:processing-of-training-example}}. Figures~\ref{img:square_duration_predict} and \ref{img:hex_duration_predict} show the predicted stress contours and the difference between NN prediction and FEA. In Figures~\ref{img:square_duration_predict}(b) and \ref{img:hex_duration_predict}(b), we observe large stress differences in StressNet prediction (in dark red and blue) while this difference is reduced when DiNN-N is used for both models. Prediction performance for 2000 samples is summarized in Supplementary Table~\ref{tab:pred_error_no_cut_2000}.

\begin{table}[h!]
\centering
\caption{Prediction error rate of 1000 samples for models without cutout regions}
\resizebox{0.9\textwidth}{!}{%
\begin{tabular}{|c|l|c|c|}
\hline
\multicolumn{2}{|c|}{\multirow{2}{*}{Prediction Error comparing to Baseline}} & SP-VR composite & HP-VR composite \\ \cline{3-4} 
\multicolumn{2}{|c|}{}                      & DiNN-N & DiNN-N \\ \hline
\multicolumn{2}{|c|}{Fiber MER reduction}   & 38\%   & 41\%   \\ \hline
\multicolumn{2}{|c|}{Matrix MER reduction}  & 20\%   & 43\%   \\ \hline
\multicolumn{2}{|c|}{Contour MSE reduction} & 21\%   & 15\%   \\ \hline
\end{tabular}%
}
\label{tab:pred_error_no_cutout}
\end{table}

\begin{figure}[h!]
\vspace{-0.1in}
	\centering
	\subfigure[]{
	\includegraphics[width=0.475\textwidth]{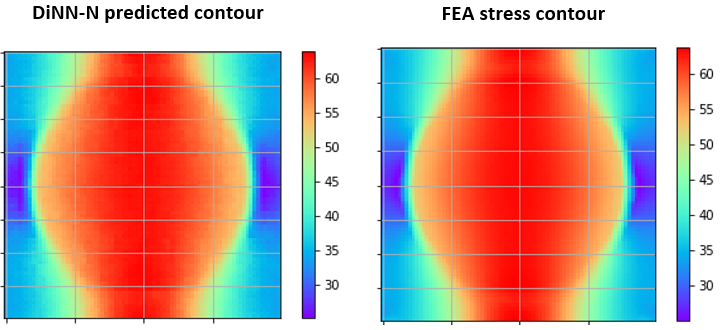} 
	}
\centering
\subfigure[]{
  \includegraphics[width=0.475\textwidth]{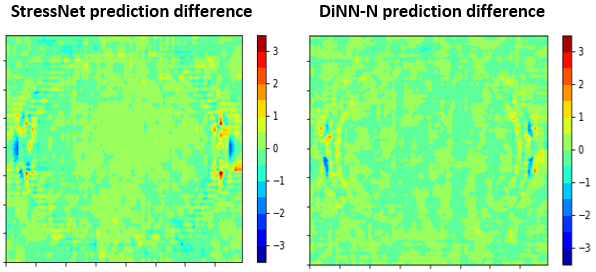}
}
\caption{SP-VR model: (a) Comparison of DiNN-N prediction and FEA stress output (b)  Comparison of difference in predicted stress distribution of StressNet and DiNN-N compared to FEA stress output}
\label{img:square_duration_predict}
\end{figure}

\begin{figure}[h!]
\vspace{-0.1in}
	\centering
	\subfigure[]{
	\includegraphics[width=0.49\textwidth]{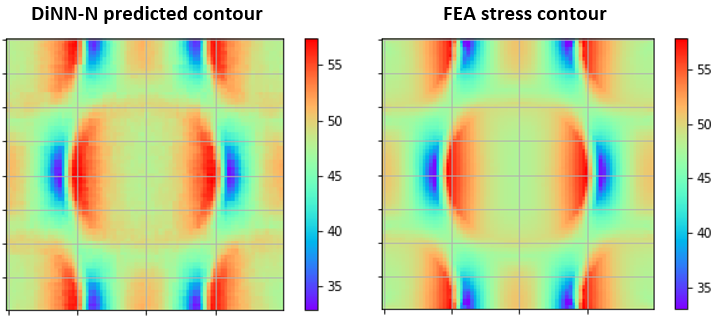}
	}
\centering
\subfigure[]{
  \includegraphics[width=0.46\textwidth]{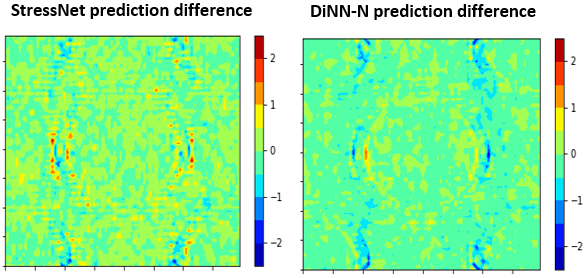}
}
\caption{HP-VR model: (a) Comparison of DiNN-N predicted and FEA stress output contours (b) Comparison of difference in predicted stress distribution of StressNet and DiNN-N compared to FEA stress output}
\label{img:hex_duration_predict}
\end{figure}

\item In general, DiNN-N significantly reduces the max stress error rate in both fiber and matrix, as well as lowers the MSE value compared to the baseline model. Hence, our proposed DiNN-N structure is shown to be a better structure for stress prediction in composite models without cutout region.

\end{itemize}

\item {\bf Prediction accuracy} for models with cutout regions:

\begin{itemize}[leftmargin=0in]
\item Table~\ref{tab:pred_error_cutout} shows the prediction performance comparing DiNN-N, DiNN-NC and baseline model StressNet for 1000 samples. DiNN-NC is proposed for models with cutout and severe stress concentration, where a clean module is added to the DiNN-N framework. The process of determining difference contours can introduce non-zero values in geometry and stress contours in regions without materials, which we refer to as noise. As noises are further magnified during training, they can adversely influence prediction in the surrounding areas, especially when prediction in these areas is important due to high stress concentrations. In DiNN-NC, the clean module removes unnecessary noises in the areas where material is absent and helps the Neural Network to improve prediction accuracy in training during the back-propagation process. Here, we show that DiNN-NC improves the performance in such cases. That is, for PC-VR composite model, DiNN-N reduces MER, however, MSE value increases compared to the baseline model. This is attributed to extra noise from cutout region during mean contour subtraction. By adding the clean module, DiNN-NC manifests the lowest MER and MSE values with significant reduction compared to the baseline model. For HPR-VR composite model, prediction accuracy of DiNN-N for maximum stress in the ring region is worse compared to baseline model due to severe stress concentration ($R_{\sigma}>$ 4) around cutout and within the ring (particle) region. The negative influence of noise in the cutout region is more significant as compared to the PC-VR model. Hence, by introducing the clean module, DiNN-NC shows higher prediction accuracy. Figure~\ref{img:hollow_duration_predict}(b) and Figure~\ref{img:particle_duration_prediction}(b) similarly show that DiNN-NC significantly reduces large prediction differences in stress distribution as highlighted by dark red and blue colors compared to baseline and DiNN-N models. Prediction performance for 2000 samples is summarized in Supplementary Table~\ref{tab:pred_error_cut_2000}.

\begin{table}[h!]
\centering
\caption{Prediction error rate of 1000 samples for models with cutout regions (red color means worse prediction comparing to baseline)}
\resizebox{0.9\textwidth}{!}{%
\begin{tabular}{|c|l|c|c|c|c|}
\hline
\multicolumn{2}{|c|}{\multirow{2}{*}{Prediction Error comparing to Baseline}} &
  \multicolumn{2}{c|}{PC-VR composite} &
  \multicolumn{2}{c|}{HP-VR composite} \\ \cline{3-6} 
\multicolumn{2}{|c|}{}                      & DiNN-N & DiNN-NC & DiNN-N & DiNN-NC \\ \hline
\multicolumn{2}{|c|}{Ring MER reduction}    & \textbf{---}    & \textbf{---}     & \textcolor{red}{-1\%}   & 39\%    \\ \hline
\multicolumn{2}{|c|}{Matrix MER reduction}  & 46\%   & 76\%    & 6\%    & 75\%    \\ \hline
\multicolumn{2}{|c|}{Contour MSE reduction} & \textcolor{red}{-105\%} & 86\%    & \textcolor{red}{-142\%} & 89\%    \\ \hline
\end{tabular}%
}
\label{tab:pred_error_cutout}
\end{table}

\begin{figure}[h!]
\vspace{-0.1in}
\centering
\subfigure[]{
\includegraphics[width=0.475\textwidth]{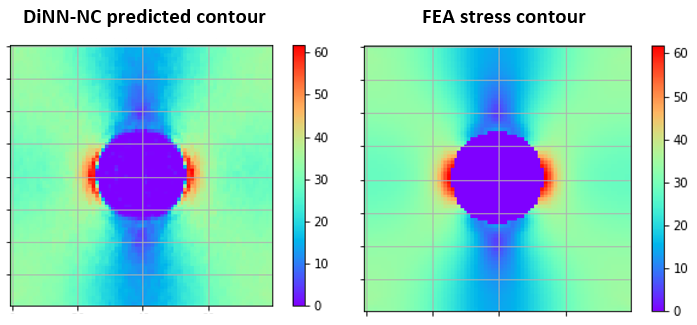} 
	}
\centering
\subfigure[]{
  \includegraphics[width=0.475\textwidth]{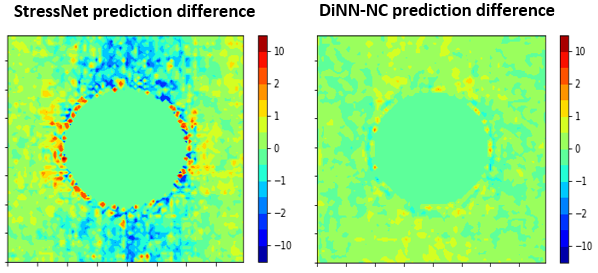}
}
\caption{PC-VR model: (a) Comparison of DiNN-NC predicted and FEA stress output contours (b) Comparison of difference in predicted stress distribution of StressNet and DiNN-NC compared to FEA stress output}
\label{img:hollow_duration_predict}
\end{figure}

\begin{figure}[h!]
\vspace{-0.1in}
	\centering
	\subfigure[]{
	\includegraphics[width=0.475\textwidth]{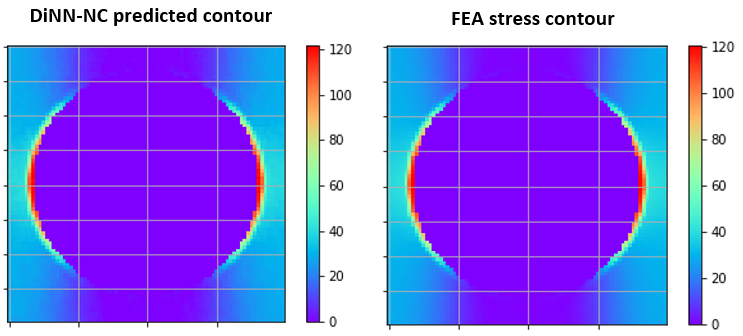}
	}
\centering
\subfigure[]{
  \includegraphics[width=0.475\textwidth]{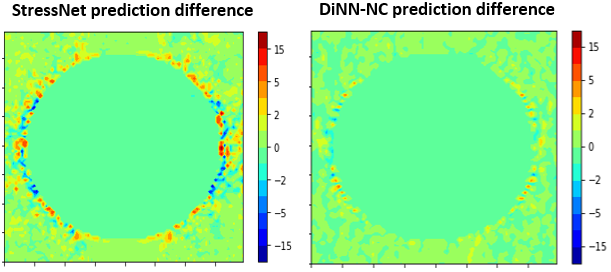}
}
\caption{HPR-VR composite: (a) Comparison of DiNN-NC predicted and FEA stress output contours (b)  Comparison of difference in predicted stress distribution of StressNet and DiNN-NC compared to FEA stress output}
\label{img:particle_duration_prediction}
\end{figure}

\item In general, we have shown that DiNN-NC structure has the best performance in terms of accurate stress prediction compared to StressNet (baseline) and DiNN-N for heterogeneous media with discontinuities, especially when significant stress concentrations exist ($R_{\sigma}>2$).

\end{itemize}

\item {\bf Prediction accuracy} for models with spatial randomness:

\begin{itemize}[leftmargin=0in]
\item Table~\ref{tab:pred_error_random} shows the prediction performance comparing DiNN-N, DiNN-NC and StressNet baseline model for test models with spatial randomness. Within this category, the performance of DiNN frameworks (DiNN-N, DiNN-NC) compared to baseline model is mixed. For SP-SR models (SP-SR-C and SP-SR-UC), DiNN-N performs better compared to baseline model except in the case of fiber MER. That is, the fiber MER is relatively larger for uncontrolled random SP-SR model. As for PC-SR models (PC-SR-C and PC-SR-UC), the prediction performance of DiNN-NC is worse in terms of MER, but better in terms of MSE, for both controlled spatial randomness and uncontrolled spatial randomness. This is attributed to the concept of reference contour representing an initial guess for stress distribution contour within our approach that gives a general guideline to the actual stress distribution contour. In that case, if our model has large spatial randomness, the reference contour cannot fully represent the geometric variations and highlight the key areas, thus posing additional challenge in predicting the difference stress contour and adversely affecting the MER. For square packed model, since no cutout or severe stress concentration exists, advantages of reference contour as initial guideline plays a positive role. Whereas, for PC-SR models, cutout regions and localized stress concentration are averaged over the reference contour, which makes the stress difference contour more unbalanced. Hence, this results in poor performance of DiNN-NC for plate with cutout models with spatial randomness which requires further analyses and is beyond the scope of this paper. Figure~\ref{img:random_control_hollow_duration_prediction}(b) and Figure~\ref{img:random_uncontrol_hollow_duration_prediction}(b) highlight the prediction difference (error) of baseline model StressNet and DiNN-NC with respect to corresponding FEA predictions shown in Figure~\ref{img:random_control_hollow_duration_prediction}(a) and Figure~\ref{img:random_uncontrol_hollow_duration_prediction}(a), respectively, corroborating that DiNN-NC has much lower prediction MSE.  

\begin{table}[h!]
\centering
\caption{Prediction error rate of 1000 samples for models with spatial randomness (red color means worse prediction comparing to baseline)}
\resizebox{0.8\textwidth}{!}{%
\begin{tabular}{|c|c|c|c|c|}
\hline
\multicolumn{2}{|c|}{\multirow{2}{*}{Prediction Error comparing to Baseline}} & \multicolumn{2}{c|}{PC-SR composite} & SP-SR composite \\ \cline{3-5} 
\multicolumn{2}{|c|}{}                                                                                     & DiNN-N & DiNN-NC & DiNN-N \\ \hline
\multirow{3}{*}{\begin{tabular}[c]{@{}c@{}}Controlled\\ randomness\end{tabular}}   & Fiber MER reduction   & \textbf{---}    & \textbf{---}     & 47\%   \\ \cline{2-5} 
                                                                                   & Matrix MER reduction  & \textcolor{red}{-72\%}  & \textcolor{red}{-23\%}   & 13\%   \\ \cline{2-5} 
                                                                                   & Contour MSE reduction & \textcolor{red}{-112\%} & 60\%    & 28\%   \\ \hline
\multirow{3}{*}{\begin{tabular}[c]{@{}c@{}}Uncontrolled\\ randomness\end{tabular}} & Fiber MER reduction   & \textbf{---}    & \textbf{---}     & \textcolor{red}{-10\%}  \\ \cline{2-5} 
                                                                                   & Matrix MER reduction  & \textcolor{red}{-173\%} & \textcolor{red}{-38\%}   & 12\%   \\ \cline{2-5} 
                                                                                   & Contour MSE reduction & \textcolor{red}{-494\%} & 45\%  & 17\%   \\ \hline
\end{tabular}%
}
\label{tab:pred_error_random}
\end{table}

\begin{figure}[h!]
\vspace{-0.1in}
	\centering
	\subfigure[]{
	\includegraphics[width=0.475\textwidth]{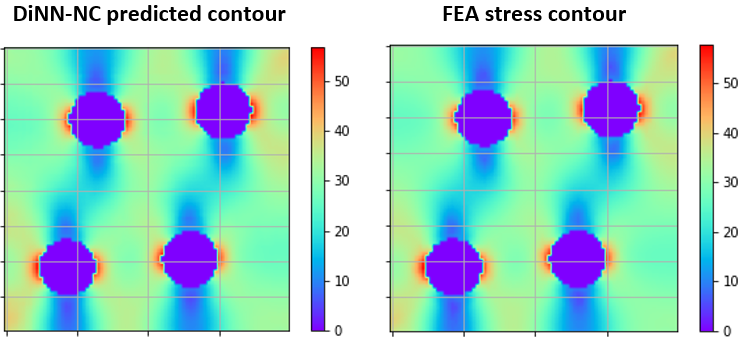}
	}
\centering
\subfigure[]{
  \includegraphics[width=0.475\textwidth]{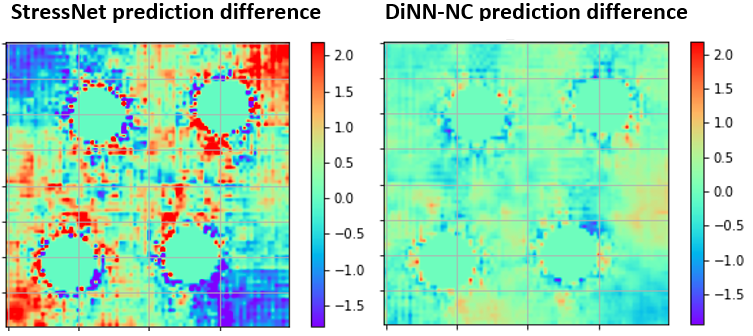}
}
\caption{PC-SR-C: (a) Comparison of DiNN-NC predicted and FEA stress output contours (b)  Comparison of difference in predicted stress distribution of StressNet and DiNN-NC compared to FEA stress output}
\label{img:random_control_hollow_duration_prediction}
\end{figure}

\begin{figure}[h!]
\vspace{-0.1in}
	\centering
	\subfigure[]{
	\includegraphics[width=0.475\textwidth]{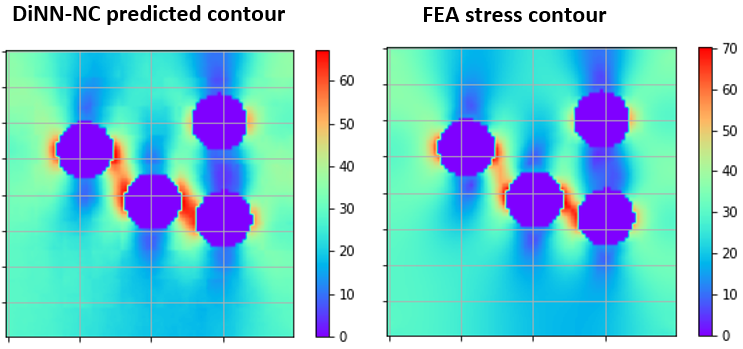}
	}
\centering
\subfigure[]{
  \includegraphics[width=0.475\textwidth]{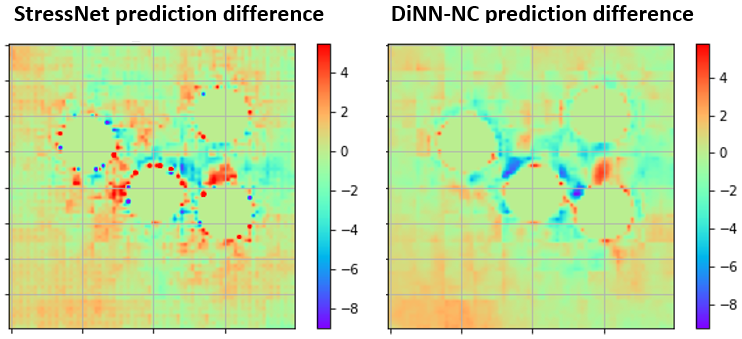}
}
\caption{PC-SR-UC: (a) Comparison of DiNN-NC predicted and FEA stress output contours (b)  Comparison of difference in predicted stress distribution of StressNet and DiNN-NC compared to FEA stress output}
\label{img:random_uncontrol_hollow_duration_prediction}
\end{figure}

\item To summarize, with limited data size, DiNN frameworks enhance MSE in their prediction for all models but could have larger MER in certain models and certain areas compared to baseline model. Further research is needed for investigating solutions to address this challenge in spatially random models.
\end{itemize}

\item {\bf Comparison of prediction duration between proposed NN and FEA}:

The goal of substituting FEA with NN is to accelerate stress prediction. \mbox{Table~\ref{tab:NN_pred_duration}} compares the time taken by traditional FEA, StressNet and DiNN frameworks for predicting stress contours within four micromechanical models. We notice that our DiNN frameworks (including DiNN-N and DiNN-NC) have similar prediction speed as the baseline StressNet model with around 0.2 seconds difference. Both these frameworks are on average about 5-6 times faster for making a single prediction compared to FEA as shown in \mbox{ Table~\ref{tab:NN_pred_duration}}. Here, we are comparing the time taken by the trained DiNN for making predictions with that taken by FEA. The observed time saving when using DiNN arises because the Neural Network predicts the distribution contours directly through a sequence of trained nonlinear equations. This is unlike in the FEA, where each model domain is discretized into a mesh and the stresses are calculated by numerically solving the weak forms of the governing equations. The NN training step happens once, and the trained model can be used for predicting stress outputs for several geometric models. This time saving is significant when used for performing micromechanical analysis in multi-scaling and optimization problems.  The prediction efficiency of DiNN will become more significant as models become increasingly complex or more design iterations are required to be tested.

\begin{table}[h!]
\centering
\caption{FEA and Neural Network prediction duration per sample}\label{tab:prediction_per_sample}
\resizebox{0.6\textwidth}{!}{%
\begin{tabular}{|c|c|c|c|c|}
\hline
           & PC-VR & SP-VR & HP-VR            & HPR-VR \\ \hline
FEA (ABAQUS)     & 6.32 sec          & 6.67 sec      & 7.50 sec                    & 8.57 sec                   \\ \hline
StressNet & 1.18 sec          & 0.96 sec      & 0.83 sec                    & 0.84 sec                   \\ \hline
DiNN        & 1.64 sec          & 1.71 sec      & 1.79 sec                    & 1.43 sec                   \\ \hline
DiNN-N      & 1.38 sec          & 1.15 sec      & 1.08 sec                    & 1.09 sec                   \\ \hline
DiNN-NC     & 1.40 sec          & 1.25 sec      & \textbf{---}                         & \textbf{---}                        \\ \hline
\end{tabular}%
}
\label{tab:NN_pred_duration}
\end{table}

\item {\bf Comparison of training duration between NN frameworks}: Baseline model StressNet in general has lower training duration compared to DiNN structures presented in this paper. This is because the DiNN structures are built upon StressNet, where the data processing module and stress prediction module within DiNN structures add additional parameters for training, consequently adding to the computational time. DiNN frameworks only have marginally longer training duration in a reasonable range. Table~\ref{tab:NN_train_duration} shows detailed information about total training duration for PC-VR and SP-VR, as an example. We observe that DiNN increases the training duration by $\approx$27\% and DiNN-NC increases the training duration by $\approx$58\%, which is around 200 seconds more for 800 samples. Among the DiNN structures presented in this paper, adding DiNN-N marginally reduces the training duration as it forces the input and output of the Encoder-Decoder module to stay within similar limits. On the other hand, adding the clean module (DiNN-NC) introduces additional computational cost. However, it should be noted here that the NN training stage is only for establishing the parameters and is a one-time event. Whereas, the prediction duration per sample for DiNN-N and DiNN-NC is on average only 0.2 seconds more compared to StressNet, as stated in part (D).

\begin{table}[h!]
\centering
\caption{Neural Network training duration for 800 samples}\label{tab:prediction_per_sample}
\resizebox{0.8\textwidth}{!}{%
\begin{tabular}{|c|l|c|c|c|c|c|}
\hline
\multicolumn{2}{|c|}{\multirow{2}{*}{\begin{tabular}[c]{@{}c@{}}Neural Network \\ Training Duration\end{tabular}}} &
  \multicolumn{3}{c|}{PC-VR composite} &
  \multicolumn{2}{c|}{SP-VR composite} \\ \cline{3-7} 
\multicolumn{2}{|c|}{}              & StressNet & DiNN-N & DiNN-NC & StressNet & DiNN-N \\ \hline
\multicolumn{2}{|c|}{Preprocessing} & 110.6 sec     & 123.7 sec & 183.9 sec  & 150.8 sec  & 148.4 sec \\ \hline
\multicolumn{2}{|c|}{Training}      & 246.0 sec     & 331.7 sec & 378.6 sec  & 198.8 sec  & 341.7 sec \\ \hline
\multicolumn{2}{|c|}{Total}         & 356.6 sec     & 455.4 sec & 562.5 sec  & 349.6 sec  & 490.1 sec \\ \hline
\end{tabular}%
}
\label{tab:NN_train_duration}
\end{table}

Based on these micromechanical models tested, we demonstrate that our DiNN framework is capable of significantly enhancing prediction accuracy compared to baseline StressNet framework, while maintaining the training duration within a reasonable range. For other new composite models, the training and prediction duration can vary depending on the model geometry and a suitable Cartesian Map size (determined by convergence study) used. However, these will affect both DiNN and baseline frameworks similarly, and thus, similar trends are expected for training and prediction duration between these frameworks.

\end{enumerate}

\vspace{-0.3in}
\section{Conclusions}

In this paper, we have presented a novel NN framework as a surrogate for traditional FEA approach to predict stress distribution in heterogeneous media, like composite materials. Our approach consists of a set of DiNN frameworks capable of predicting stress distributions with very high accuracy for different types of composite materials. Two main types of composites are considered: models with volume fraction randomness and models with spatial randomness. Eight different composite micromechanical models were considered for validating the performance of our NN structures. For models with volume fraction randomness, the proposed DiNN structure included a normalization module (DiNN-N) for all geometries considered, while we additionally introduced a clean module with DiNN-N, named DiNN-NC, for geometries with discontinuities. 
For models with spatial randomness, DiNN frameworks perform well in terms of prediction MSE, but can have mixed performance in terms of MER. Further research effort is needed to resolve unstable MER for spatially random distributed models.

The DiNN frameworks presented in this paper for stress prediction in heterogeneous media can be used in future studies including mechanical property prediction and composite structure optimization at multiple length scales.

\noindent Key contributions of this paper are:

\begin{enumerate}[leftmargin=0.2in,noitemsep]
	\item This is the first attempt to our best knowledge that brings Convolutional Neural Network based Machine Learning for stress distribution prediction for heterogeneous media like composite materials considering volume fraction and spatial randomness.
	\item We introduce a novel {\bf Difference-based} Neural Network framework which utilizes a set of reference models from the training set and focuses especially on training the difference contours between the target model and the reference set. This is shown to improve stress prediction accuracy when high stress concentrations manifest within heterogeneous media, with or without discontinuities like cutouts.
	\item We show that the Difference-based Neural Network framework improves the stress prediction accuracy significantly compared to existing baseline structures, especially on models with volume fraction randomness and large local stress concentrations. For models with spatial randomness, our framework guarantees a reliable prediction in terms of Mean-Squared-Error. Moreover, our proposed framework has a faster prediction speed compared to traditional Finite Element Method.
\end{enumerate}

\section*{Acknowledgements}
The authors would like to acknowledge the support through the 2019 ONR Young Investigator Program [N00014-19-1-2206] through {\em{Sea-based Aviation: Structures and Materials Program}} for conducting the research presented here. The authors would also like to acknowledge the support from the University of Wisconsin Graduate Fellowship.


\section*{Data Availability}
The entire DiNN framework can be found on our GitHub page: \url{https://github.com/Isaac0047/DiNN-framework-Stress-Prediction-for-Heterogeneous-Media.git}. This includes the entire implemented code with model generation, data processing and DiNN framework setup. A readme file describes the steps involved with running the DiNN framework. Also, details about how to run the code with an example is provided on this GitHub page. The raw data required to reproduce the findings presented in the paper are available to download from \url{https://drive.google.com/drive/folders/1jTvtSSgwK-qvONTyDTAzZZFvRUcZNQ-R?usp=sharing}. 


{\footnotesize
\bibliographystyle{unsrt}
\bibliography{sample}

\begin{thebibliography}{10}

\bibitem{The-FEM}
T.~Hughes.
\newblock {\em The Finite Element Method: Linear Static and Dynamic Finite
  Element Analysis}, volume~78.
\newblock 01 2000.

\bibitem{Nonlinear-FEM}
P.~Wriggers.
\newblock {\em Nonlinear Finite Element Methods}, volume~4.
\newblock 01 2008.

\bibitem{fem-composite-book}
E.~Barbero.
\newblock {\em Finite Element Analysis of Composite Materials}.
\newblock 01 2008.

\bibitem{EFENDIEV2013116}
Y.~Efendiev, J.~Galvis, and T.~Y. Hou.
\newblock Generalized multiscale finite element methods (gmsfem).
\newblock {\em Journal of Computational Physics}, 251:116 -- 135, 2013.

\bibitem{HOU1997169}
T.~Y. Hou and X.~Wu.
\newblock A multiscale finite element method for elliptic problems in composite
  materials and porous media.
\newblock {\em Journal of Computational Physics}, 134(1):169 -- 189, 1997.

\bibitem{FEA-composite}
E.~Barbero.
\newblock {\em Finite Element Analysis of Composite Materials}.
\newblock 01 2008.

\bibitem{3D_autoencoder}
N.~Umetani.
\newblock Exploring generative 3d shapes using autoencoder networks.
\newblock pages 1--4, 11 2017.

\bibitem{Data_driven_topology}
E.~Ulu, R.~Zhang, and L.~Kara.
\newblock A data-driven investigation and estimation of optimal topologies
  under variable loading configurations.
\newblock {\em Computer Methods in Biomechanics and Biomedical Engineering:
  Imaging \& Visualization}, 4:1--12, 08 2015.

\bibitem{VAMPnets}
A.~Mardt, L.~Pasquali, H.~Wu, and F.~Noé.
\newblock Vampnets: Deep learning of molecular kinetics.
\newblock {\em Nature Communications}, 9, 10 2017.

\bibitem{Self_Learn_molecular}
G.~Tribello, M.~Ceriotti, and M.~Parrinello.
\newblock A self-learning algorithm for biased molecular dynamics.
\newblock {\em Proceedings of the National Academy of Sciences of the United
  States of America}, 107:17509--17514, 01 2010.

\bibitem{11-Nourbak}
J.~Irizarry M.~Nourbakhsh and J.~Haymaker.
\newblock Generalizable surrogate model features to approxi-mate stress in 3d
  trusses.
\newblock {\em Engineering Applications of Artiﬁcial Intelligence}, pages
  15--27, 01 2018.

\bibitem{12-Khadilkar}
J.~Wang A.~Khadilkar and R.~Rai.
\newblock Deep learning–based stress prediction for bottom-up sla 3d printing
  process.
\newblock {\em The International Journal of Ad-vanced Manufacturing
  Technology}, pages 1--15, 2019.

\bibitem{14-KHOZANI2017441}
Z.~S. Khozani, H.~Bonakdari, and A.~H. Zaji.
\newblock Estimating the shear stress distribution in circular channels based
  on the randomized neural network technique.
\newblock {\em Applied Soft Computing}, 58:441 -- 448, 2017.

\bibitem{16-DBLP:journals/corr/IsolaZZE16}
P.~Isola, J.~Y. Zhu, T.~Zhou, and A.~A. Efros.
\newblock Image-to-image translation with conditional adversarial networks.
\newblock {\em CoRR}, abs/1611.07004, 2016.

\bibitem{10-LiangDL}
C.~Martin L.~Liang and S.~Sun.
\newblock A deep learning approach to estimate stress distribution: a fast and
  accurate surrogate of ﬁnite-element analysis.
\newblock {\em Journal of The Royal Society Interface, 15, 01}, 01 2018.

\bibitem{10.1145/2939672.2939738}
X.~Guo, W.~Li, and F.~Iorio.
\newblock Convolutional neural networks for steady flow approximation.
\newblock In {\em Proceedings of the 22nd ACM SIGKDD International Conference
  on Knowledge Discovery and Data Mining}, KDD ’16, page 481–490, New York,
  NY, USA, 2016. Association for Computing Machinery.

\bibitem{Encoder-Decoder}
K.~Cho, v.~M.~Bart, D.~Bahdanau, and Y.~Bengio.
\newblock On the properties of neural machine translation: Encoder-decoder
  approaches.
\newblock 09 2014.

\bibitem{21-ATALLA1998135}
M.~J. Atalla and D.~J. Inman.
\newblock On model updating using neural networks.
\newblock {\em Mechanical Systems and Signal Processing}, 12(1):135 -- 161,
  1998.

\bibitem{22-inproceedings}
T.~Schröppel T.~Spruegel and S.~Wartzack.
\newblock Generic approach to plausibility checks for structural mechanics with
  deep learning.
\newblock 08 2017.

\bibitem{24-article}
A.~Oishi and G.~Yagawa.
\newblock Computational mechanics enhanced by deep learning.
\newblock {\em Computer Methods in Applied Mechanics and Engineering}, 327, 09
  2017.

\bibitem{Afshar-CFD}
Y.~Afshar, S.~Bhatnagar, S.~Pan, K.~Duraisamy, and S.~Kaushik.
\newblock Prediction of aerodynamic flow fields using convolutional neural
  networks.
\newblock 05 2019.

\bibitem{Nie-Stress-Net}
Z.~Nie, H.~Jiang, and L.~Kara.
\newblock Stress field prediction in cantilevered structures using
  convolutional neural networks.
\newblock {\em Journal of Computing and Information Science in Engineering},
  page~1, 06 2019.

\bibitem{He-ResNet}
K.~He, X.~Zhang, S.~Ren, and J.~Sun.
\newblock Deep residual learning for image recognition.
\newblock pages 770--778, 06 2016.

\bibitem{article}
Y.~Bengio and Y.~Lecun.
\newblock Convolutional networks for images, speech, and time-series.
\newblock 11 1997.

\bibitem{seq2seq}
I.~Sutskever, O.~Vinyals, and Q.~Le.
\newblock Sequence to sequence learning with neural networks.
\newblock {\em Advances in Neural Information Processing Systems}, 4, 09 2014.

\bibitem{ABAQUS}
ABAQUS (2011)~Dassualt Systèmes.
\newblock Abaqus documentation.
\newblock 2011.

\bibitem{Breunig2020}
P.~Breunig, V.~Damodaran, K.~Shahapurkar, S.~Waddar, M.~Doddamani, P.~Jeyaraj,
  and P.~Prabhakar.
\newblock Dynamic impact behavior of syntactic foam core sandwich composites.
\newblock {\em Journal of Composite Materials}, 54(4):535--547, 2020.

\bibitem{Shahapurkar2018}
K.~Shahapurkar, C.~D. Garcia, M.~Doddamani, G.~C.~M. Kumar, and P.~Prabhakar.
\newblock {Compressive behavior of cenosphere/epoxy syntactic foams in arctic
  conditions}.
\newblock {\em Composites Part B: Engineering}, 135(June 2017):253--262, 2018.

\bibitem{GARCIA2018}
C.~D. Garcia, K.~Shahapurkar, M.~Doddamani, G.~C.~M. Kumar, and P.~Prabhakar.
\newblock {Effect of arctic environment on flexural behavior of fly ash
  cenosphere reinforced epoxy syntactic foams}.
\newblock {\em Composites Part B: Engineering}, 151:265--273, 2018.

\bibitem{Dodda_Microballoon}
M.~L. Jayavardhan and D.~Mrityunjay.
\newblock Quasi-static compressive response of compression molded glass
  microballoon/hdpe syntactic foam.
\newblock {\em Composites Part B Engineering}, 149, 06 2018.

\bibitem{Jones1975MechanicsOC}
R.~M. Jones.
\newblock Mechanics of composite materials, second edition.
\newblock CRC Press, 1999.

\bibitem{PRABHAKAR2013-1}
P.~Prabhakar and A.~M. Waas.
\newblock Interaction between kinking and splitting in the compressive failure
  of unidirectional fiber reinforced laminated composites.
\newblock {\em Composite Structures}, 98:85 -- 92, 2013.

\bibitem{PRABHAKAR2013-2}
P.~Prabhakar and A.~M. Waas.
\newblock Micromechanical modeling to determine the compressive strength and
  failure mode interaction of multidirectional laminates.
\newblock {\em Composites Part A: Applied Science and Manufacturing}, 50:11 --
  21, 2013.

\bibitem{PRABHAKAR2013-4}
P.~Prabhakar and A.~M. Waas.
\newblock Upscaling from a micro-mechanics model to capture laminate
  compressive strength due to kink banding instability.
\newblock {\em Computational Materials Science}, 67:40 -- 47, 2013.

\bibitem{High_E_CF_stanage}
A.~E. Standage and R.~Prescott.
\newblock High elastic modulus carbon fibre.
\newblock {\em Nature}, page 211, 07 1966.

\bibitem{Min-Max}
P.~Juszczak, D.~Tax, and R.~Duin.
\newblock Feature scaling in support vector data description.
\newblock 05 2002.

\bibitem{hu2018senet}
J.~Hu, L.~Shen, and G.~Sun.
\newblock Squeeze-and-excitation networks.
\newblock 2018.

\bibitem{NIPS2014_5485}
L.~Xu, J.~Ren, C.~Liu, and J.~Jia.
\newblock Deep convolutional neural network for image deconvolution.
\newblock In Z.~Ghahramani, M.~Welling, C.~Cortes, N.~D. Lawrence, and K.~Q.
  Weinberger, editors, {\em Advances in Neural Information Processing Systems
  27}, pages 1790--1798. Curran Associates, Inc., 2014.

\bibitem{Glorot10understandingthe}
G.~Xavier and B.~Yoshua.
\newblock Understanding the difficulty of training deep feedforward neural
  networks.
\newblock In {\em In Proceedings of the International Conference on Artificial
  Intelligence and Statistics (AISTATS’10). Society for Artificial
  Intelligence and Statistics}, 2010.

\bibitem{Hadamard}
G.~P.~H. Styan.
\newblock Hadamard products and multivariate statistical analysis.
\newblock {\em Linear Algebra and its Applications}, 6:217 -- 240, 1973.

\bibitem{Tri-interp-Amidror}
I.~Amidror.
\newblock Scattered data interpolation methods for electronic imaging systems:
  A survey.
\newblock {\em Journal of Electronic Imaging}, 11, 04 2002.

\bibitem{Barycentric}
E.~Hille.
\newblock {\em Analytic Function Theory}, volume~1.
\newblock Chelsea Publishing Company, New York, 2 edition, 1982.
\newblock footnote 1.

\bibitem{math-stat}
J.~F. Kenney and E.~S. Keeping, editors.
\newblock {\em Mathematics of Statistics}.
\newblock second edition. Van Nostrand, New York, 1951.

\bibitem{class_imbalance}
N.~Japkowicz and S.~Shaju.
\newblock The class imbalance problem: A systematic study.
\newblock {\em Intell. Data Anal.}, 6:429--449, 11 2002.

\bibitem{DBLP:journals/corr/abs-1305-1707}
R.~Longadge and S.~Dongre.
\newblock Class imbalance problem in data mining review.
\newblock {\em CoRR}, abs/1305.1707, 2013.

\bibitem{SUBBANARASIMHA2000117}
P.~N. SubbaNarasimha, B.~Arinze, and M.~Anandarajan.
\newblock The predictive accuracy of artificial neural networks and multiple
  regression in the case of skewed data: exploration of some issues.
\newblock {\em Expert Systems with Applications}, 19(2):117 -- 123, 2000.

\bibitem{skew-transform}
J.~D. Emerson, editor.
\newblock {\em Mathematical aspects of transformation}, pages 247--282.
\newblock John Wiley, New York, 1983.

\bibitem{pseudo-number}
P.~L’Ecuyer.
\newblock {\em Pseudorandom Number Generators}.
\newblock 05 2010.

\bibitem{scikit-learn}
F.~Pedregosa, G.~Varoquaux, A.~Gramfort, V.~Michel, B.~Thirion, O.~Grisel,
  M.~Blondel, P.~Prettenhofer, R.~Weiss, V.~Dubourg, J.~Vanderplas, A.~Passos,
  D.~Cournapeau, M.~Brucher, M.~Perrot, and E.~Duchesnay.
\newblock Scikit-learn: Machine learning in {P}ython.
\newblock {\em Journal of Machine Learning Research}, 12:2825--2830, 2011.

\bibitem{bottou_1999}
L.~Bottou.
\newblock {\em On-line Learning and Stochastic Approximations}, page 9–42.
\newblock Publications of the Newton Institute. Cambridge University Press,
  1999.

\end{thebibliography}
}


\newpage

\section*{Appendix}
\addcontentsline{toc}{section}{Appendices}
\renewcommand{\thesubsection}{\Alph{subsection}}
\renewcommand{\thefigure}{\Alph{subsection}.\arabic{figure}}
\setcounter{figure}{0}

\subsection{Brief overview of Finite Element Analysis}\label{FEM}

To obtain the physical property contours, the target solid domain is first discretized by meshing with different types of elements. After assigning material properties, boundary conditions and loads, this meshed domain is passed into a FEA solver. The FEA solver first calculates element stiffness matrices and force vectors based on individual element as shown in Equation~\ref{eqn:element_K} and \ref{eqn:element_F} as:

\begin{equation}
    K^e=\int_{V_{e}} B_{e}^T \mathbb{C}_{e} B_{e} dV_e
\label{eqn:element_K}
\end{equation}

\begin{equation}
    P^e=\int_{V_{e}} N^T f dV_e + \int_{S_{e}} N^T t dS_e
\label{eqn:element_F}
\end{equation}

\noindent where, N is the shape function, B is the strain-displacement matrix based on shape function and C is the constitutive matrix defined with elastic modulus and Poisson's ratio. f represents the body force and t represents external traction.

The element stiffness matrices $K^e$ and force vectors $P^e$ are assembled over all elements in meshed geometry into global stiffness matrix $K$ and force vector $P$, and solved for the nodal displacements $u$ as well as stresses $\sigma$ using Equation~\ref{eqn:global_eqn} and \ref{eqn:sigma} for static state analysis. Upon extracting the nodal stresses, the 2D von Mises stress can be further calculated as defined in Equation~\ref{eqn:von}.

\begin{equation}
    Ku=P
\label{eqn:global_eqn}
\end{equation}

\begin{equation}
    \sigma=CBu
\label{eqn:sigma}
\end{equation}

\begin{equation}
    \sigma_{von Mises}=\sqrt{{\sigma_{x}}^2+{\sigma_{y}}^2-\sigma_{x}\sigma_{y}+3{\tau_{xy}}^2}
\label{eqn:von}
\end{equation}

\subsection{Impact of skewness}\label{sse:skewness}
\renewcommand{\figurename}{Supplementary Figure}

\begin{figure}[h!]
	\centering
	\includegraphics[width=0.7\textwidth]{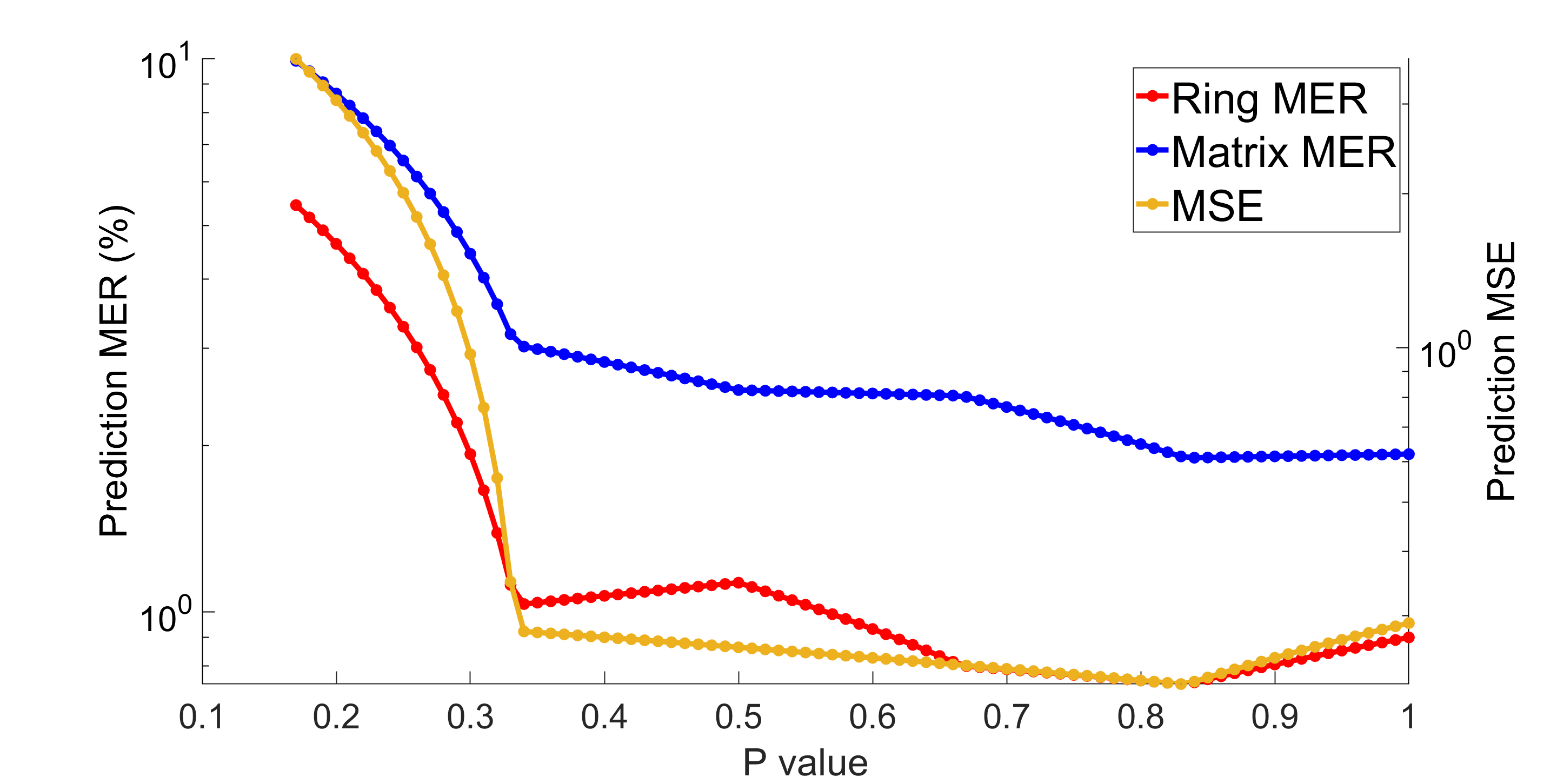} 
	\caption{Influence of skewness power $p$ on prediction error}
	\label{img:p_accuracy}
\end{figure}

To understand the effect of skewness in training samples while training the DiNN and to determine the optimal value for the power $p$ in Equation~\ref{eqn:sigma_p} as discussed in Section~\ref{subsec:analyze-skewness}, HPR-VR model is chosen as it manifests most severe stress concentrations compared to other models considered in this paper. We investigated the relationship between the accuracy of the NN prediction and $p$ values ranging between 0 and 1. Figure~\ref{img:p_accuracy} shows the variation of predicted MER in the ring and matrix regions as well as MSE for different values of $p$. The results from this analysis show that the prediction error increases rapidly when $p$ < 0.35 and is the lowest when $p$ = 0.84. However, the improvement in prediction accuracy is marginal with $p$ = 0.84 when compared to $p$ = 1. That is, the DiNN structure can effectively reduce the effect of skewness on stress contour prediction without considering any skewness reduction method, hence, a $p$ value of 1 is suggested for convenience.

\subsection{Figures}
\setcounter{figure}{0}
\renewcommand{\figurename}{Supplementary Figure}

\begin{figure}[H]
	\centering
	\includegraphics[width=0.6\textwidth]{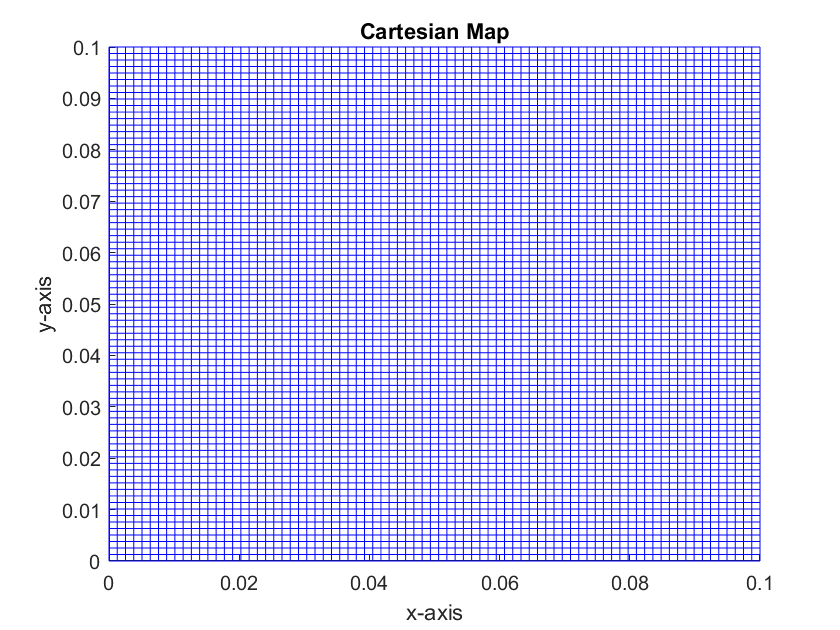}
	\caption{80-by-80 Cartesian Map matrix}
	\label{img:CM}
\end{figure}

\begin{figure}[H]
	\centering
	\includegraphics[width=0.8\textwidth]{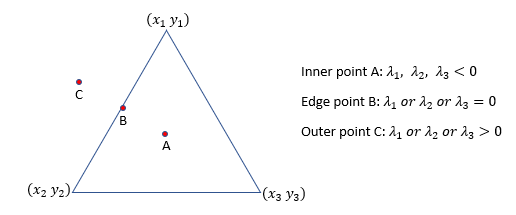}
	\caption{Relative position between target node (A,B,C) of the Cartesian Map and triangular element from the Finite Element mesh}
	\label{img:Rela_pos}
	
\end{figure}

\begin{figure}[H]
	\centering
	\includegraphics[width=\textwidth]{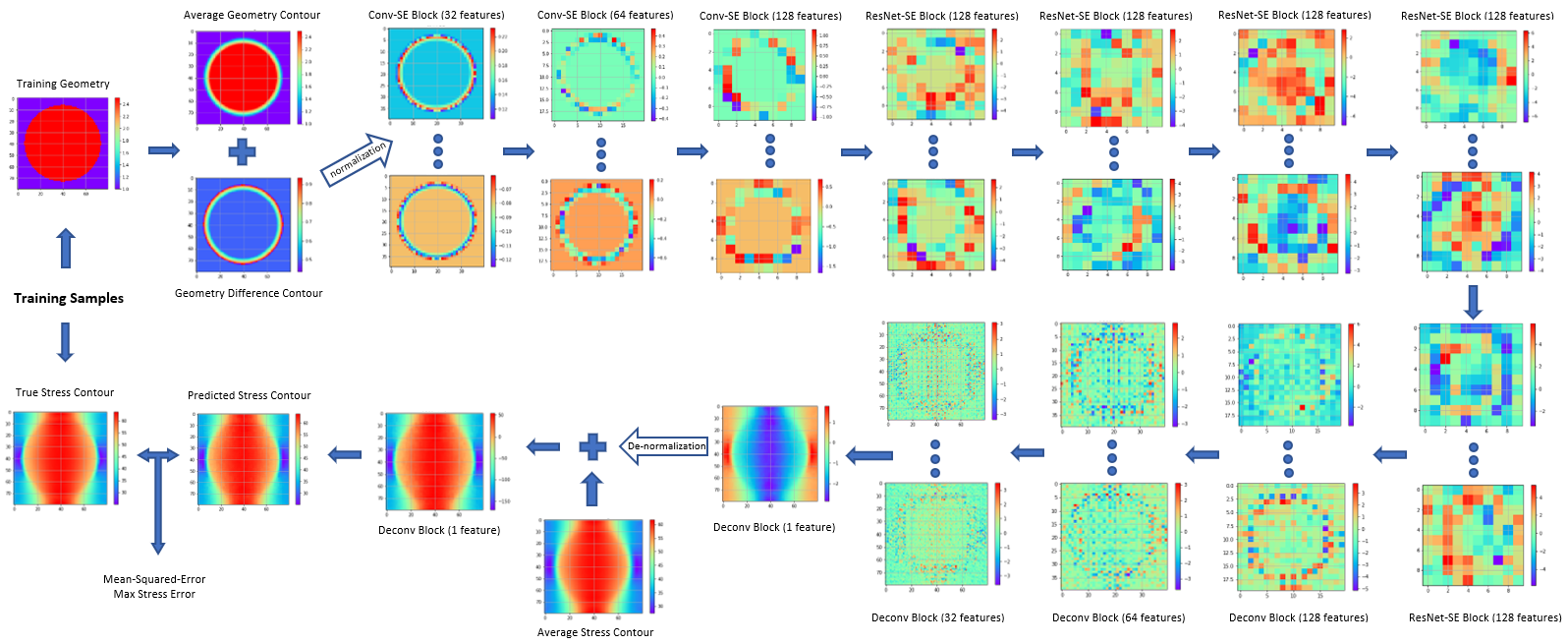}
	\caption{An examples of a Difference-based Neural Network prediction flowchart on square packed fiber reinforced model}
	\label{img:NN-flowchart}
\end{figure}

\begin{figure}[H]
\centering
\subfigure[]{
  \includegraphics[width=0.47\textwidth]{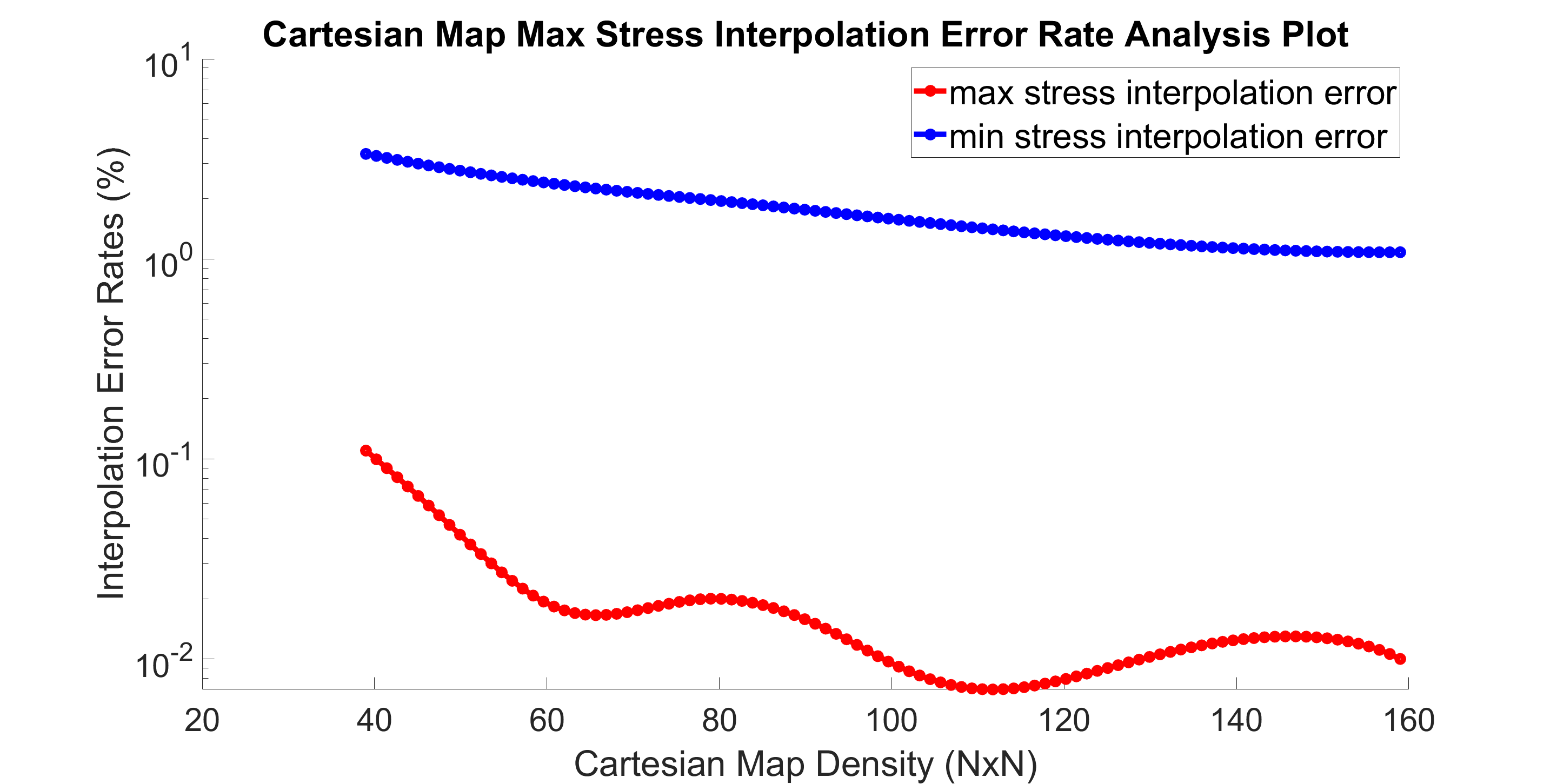}
}
\centering
\subfigure[]{
  \includegraphics[width=0.47\textwidth]{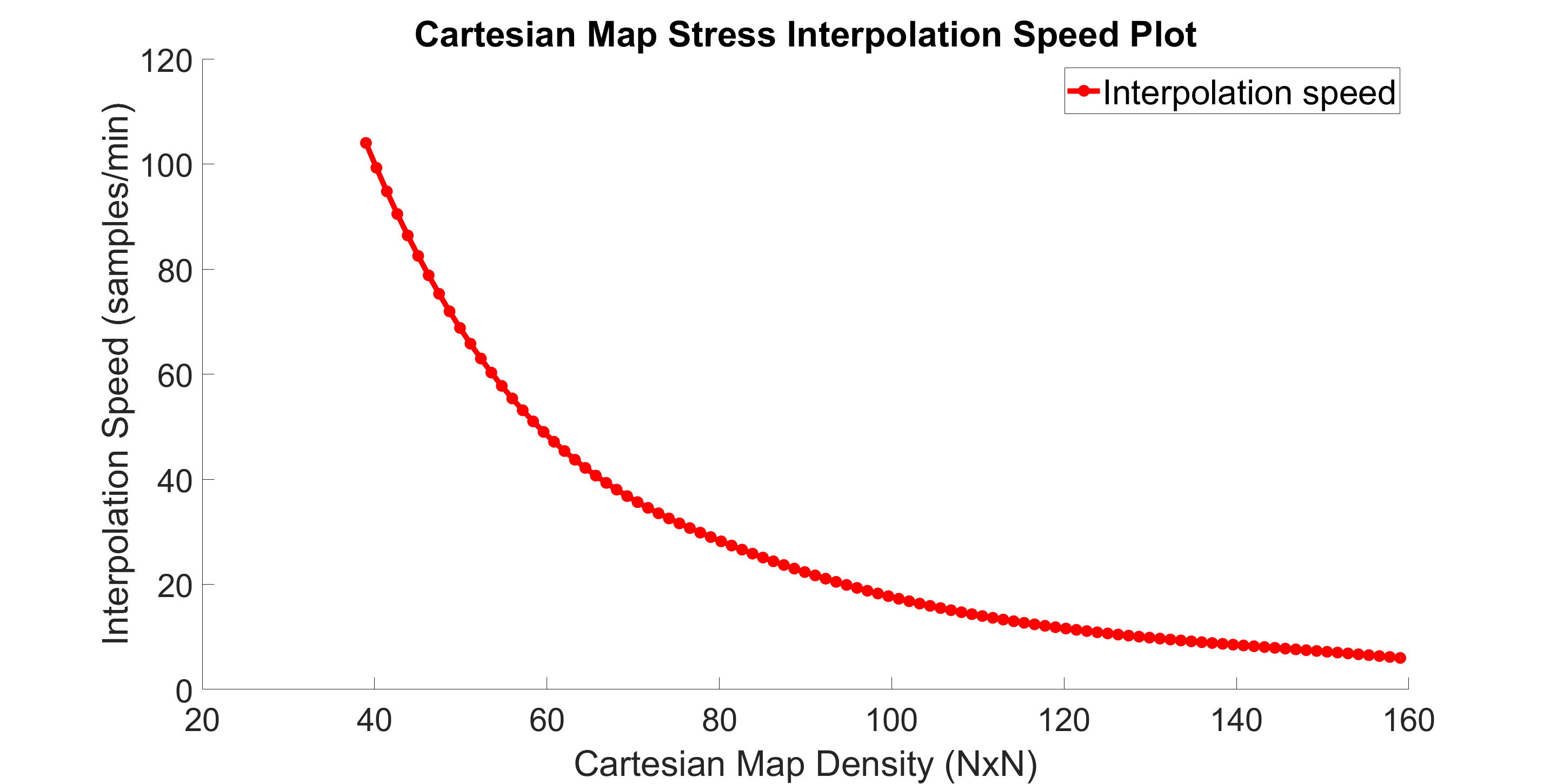}
}
\caption{(a) Cartesian Map interpolation error rate (b) Cartesian Map interpolation speed}
\label{img:CM_interp}
\end{figure}

\begin{figure}[H]
\centering
\subfigure[]{
  \includegraphics[width=0.21\textwidth]{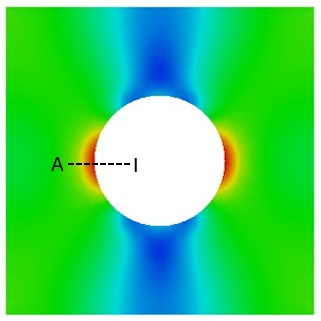}
}
\centering
\subfigure[]{
  \includegraphics[width=0.21\textwidth]{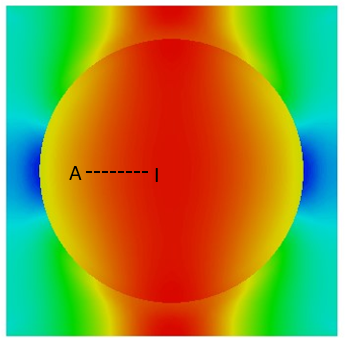}
}
\centering
\subfigure[]{
  \includegraphics[width=0.21\textwidth]{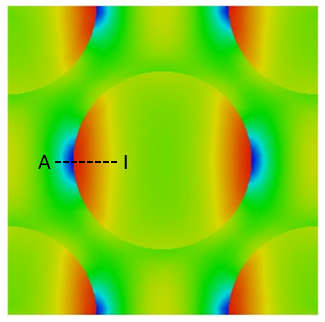}
}
\centering
\subfigure[]{
  \includegraphics[width=0.21\textwidth]{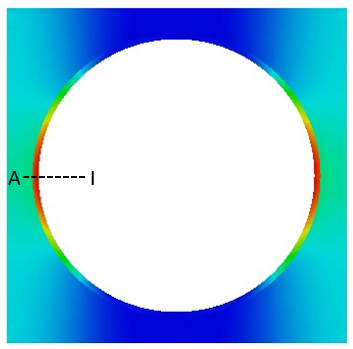}
}
\caption{Points selected for statistical analysis in (a) PC-VR (b) SP-VR (c) HP-VR (d) HPR-VR}
\label{img:statistic-point}
\end{figure}

\begin{figure}[H]
\centering
\subfigure[]{
  \includegraphics[width=0.4\textwidth]{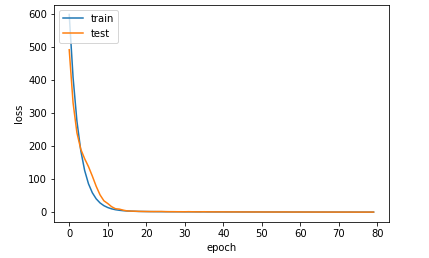}
}
\centering
\subfigure[]{
  \includegraphics[width=0.4\textwidth]{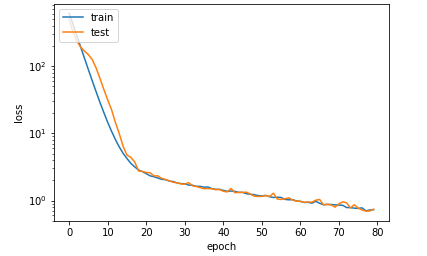}
}
\centering
\subfigure[]{
  \includegraphics[width=0.4\textwidth]{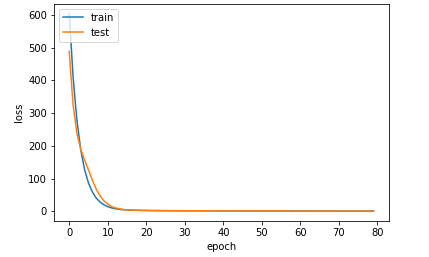}
}
\centering
\subfigure[]{
  \includegraphics[width=0.4\textwidth]{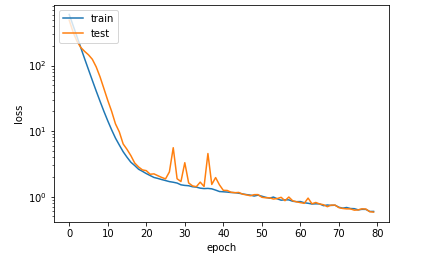}
}
\caption{Training loss for PC-VR with: (a) 1000 samples in linear scale (b) 1000 samples in log scale (c) 2000 samples in linear scale (d) 2000 samples in log scale}
\label{img:hollow_loss}
\end{figure}

\begin{figure}[H]
\centering
\subfigure[]{
  \includegraphics[width=0.42\textwidth]{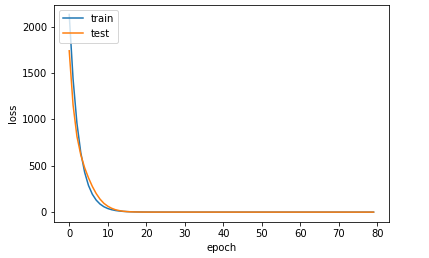}
}
\centering
\subfigure[]{
  \includegraphics[width=0.42\textwidth]{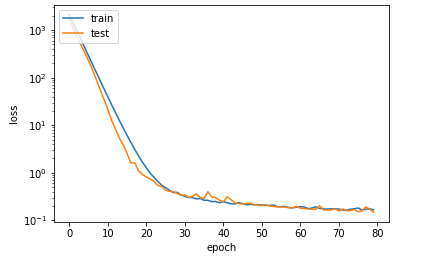}
}
\centering
\subfigure[]{
  \includegraphics[width=0.42\textwidth]{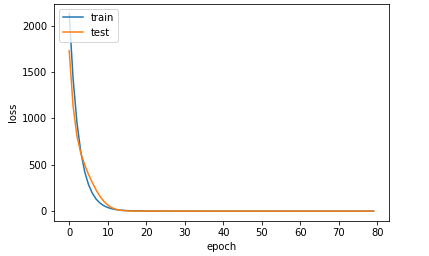}
}
\centering
\subfigure[]{
  \includegraphics[width=0.42\textwidth]{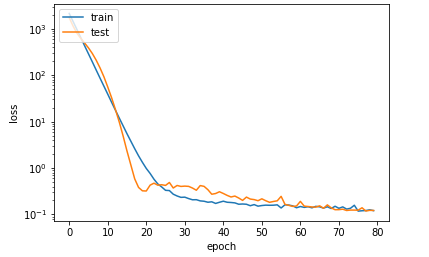}
}
\caption{Training loss for SP-VR with: (a) 1000 samples in linear scale (b) 1000 samples in log scale (c) 2000 samples in linear scale (d) 2000 samples in log scale}
\label{img:square_loss}
\end{figure}

\subsection{Tables}
\renewcommand{\thetable}{D.\arabic{table}}
\setcounter{table}{0}
\renewcommand{\tablename}{Supplementary Table}

\begin{table}[H]
\caption {Max and min values of model samples (DiNN/N = DiNN \& DiNN-N)}
\resizebox{\textwidth}{!}{%
\begin{tabular}{|c|c|c|c|c|c|c|c|c|}
\hline
\multirow{2}{*}{} & \multicolumn{3}{c|}{PC-VR} & \multicolumn{2}{c|}{SP-VR} & \multicolumn{3}{c|}{PC-SR-UC} \\ \cline{2-9} 
                   & StressNet & DiNN/N & DiNN-NC & StressNet & DiNN/N & StressNet & DiNN/N & DiNN-NC \\ \hline
Geometry max label & 1          & 0.95        & 0.95    & 1          & 0.99        & 1          & 0.44        & 0.44    \\ \hline
Geometry min label & 0          & -0.95       & 0       & 0          & -0.99       & 0          & -1          & 0       \\ \hline
Stress max value   & 68.21      & 64.80       & 64.80   & 67.83      & 14.85       & 85.89      & 65.80       & 65.80   \\ \hline
Stress min value   & 0          & -42.03      & -15.21  & 23.29      & -16.11      & 0          & -33.87      & -22.37  \\ \hline
\end{tabular}%
}
\label{tab:max-min}
\end{table}

\begin{table}[H]
\centering
\caption{Statistics properties of 1000 samples for PC-VR composite}
\resizebox{0.85\textwidth}{!}{%
\begin{tabular}{|c|c|c|c|c|c|c|c|c|c|}
\hline
                           & \multicolumn{3}{c|}{StressNet}               & \multicolumn{3}{c|}{DiNN / DiNN-N}            & \multicolumn{3}{c|}{DiNN-NC}          \\ \cline{2-10} 
\multirow{-2}{*}{Point \#} & Mean  & Median & Skewness                     & Mean & Median & Skewness                     & Mean  & Median & Skewness                    \\ \hline
A                          & 35.25 & 34.30  & {0.63}  & 0.00 & -0.95  & {0.63}  & 0.00  & -0.95  & {0.63} \\ \hline
B                          & 36.96 & 35.80  & {0.61}  & 0.00 & -1.15  & {0.61}  & 0.00  & -1.15  & {0.61} \\ \hline
C                          & 39.01 & 37.79  & {0.54}  & 0.00 & -1.22  & {0.54}  & 0.00  & -1.22  & {0.54} \\ \hline
D                          & 41.34 & 40.42  & {0.35}  & 0.00 & -0.92  & {0.35}  & 0.00  & -0.92  & {0.35} \\ \hline
E                          & 38.93 & 39.98  & {-0.22} & 0.00 & 1.05   & {-0.22} & 3.43  & 1.05   & {0.89} \\ \hline
F                          & 35.34 & 39.82  & {-0.71} & 0.00 & 4.48   & {-0.71} & 6.82  & 4.48   & {0.85} \\ \hline
G                          & 32.23 & 38.76  & {-0.90} & 0.00 & 6.53   & {-0.90} & 9.25  & 6.53   & {0.87} \\ \hline
H                          & 28.17 & 36.26  & {-1.03} & 0.00 & 8.09   & {-1.03} & 11.07 & 8.09   & {0.80} \\ \hline
I                          & 24.20 & 34.31  & {-1.23} & 0.00 & 10.11  & -1.23   & 11.93 & 10.11  & {0.41} \\ \hline
\end{tabular}%
}
\label{tab:plate-cut-stat}
\end{table}

\begin{table}[H]
\centering
\caption{Statistics properties of 1000 samples for SP-VR composite}
\resizebox{0.6\textwidth}{!}{%
\begin{tabular}{|c|c|c|c|c|c|c|}
\hline
  & \multicolumn{3}{c|}{StressNet}             & \multicolumn{3}{c|}{DiNN / DiNN-N}          \\ \cline{2-7} 
\multirow{-2}{*}{Point \#} & \multicolumn{1}{c|}{Mean} & \multicolumn{1}{c|}{Median} & Skewness & \multicolumn{1}{c|}{Mean} & \multicolumn{1}{c|}{Median} & Skewness \\ \hline
A & 58.39 & 58.33 & 0.07 & 0.00 & -0.06 & 0.07 \\ \hline
B & 58.72 & 58.66 & 0.07 & 0.00 & -0.06 & 0.07 \\ \hline
C & 59.04 & 58.98 & 0.07 & 0.00 & -0.06 & 0.07 \\ \hline
D & 59.34 & 59.27 & 0.07 & 0.00 & -0.06 & 0.07 \\ \hline
E & 59.62 & 59.55 & 0.07 & 0.00 & -0.07 & 0.07 \\ \hline
F & 59.87 & 59.82 & 0.06 & 0.00 & -0.06 & 0.06 \\ \hline
G & 60.11 & 60.05 & 0.06 & 0.00 & -0.06 & 0.06 \\ \hline
H & 60.33 & 60.27 & 0.07 & 0.00 & -0.07 & 0.07 \\ \hline
I & 60.53 & 60.47 & 0.06 & 0.00 & -0.06 & 0.06 \\ \hline
\end{tabular}%
}
\label{tab:square-stat}
\end{table}

\begin{table}[H]
\centering
\caption{Statistics properties of 1000 samples for plate with PC-SR-UC composite}
\resizebox{0.9\textwidth}{!}{%
\begin{tabular}{|c|c|c|c|c|c|c|c|c|c|}
\hline
\multirow{2}{*}{Point \#} & \multicolumn{3}{c|}{StressNet} & \multicolumn{3}{c|}{DiNN/DiNN-N} & \multicolumn{3}{c|}{DiNN-NC} \\ \cline{2-10} 
                          & Mean    & Median   & Skewness   & Mean    & Median    & Skewness   & Mean   & Median  & Skewness  \\ \hline
A & 33.25 & 32.81 & 0.49  & 0.00 & -0.44 & 0.49  & 0.00 & -0.44 & 0.49 \\ \hline
B & 33.40 & 32.98 & 0.42  & 0.00 & -0.43 & 0.42  & 0.00 & -0.43 & 0.42 \\ \hline
C & 33.57 & 33.02 & 0.47  & 0.00 & -0.55 & 0.47  & 0.00 & -0.55 & 0.47 \\ \hline
D & 33.75 & 33.06 & 0.48  & 0.00 & -0.68 & 0.48  & 0.00 & -0.68 & 0.48 \\ \hline
E & 33.92 & 32.98 & 0.53  & 0.00 & -0.95 & 0.53  & 0.00 & -0.95 & 0.53 \\ \hline
F & 33.33 & 32.80 & 0.24  & 0.00 & -0.54 & 0.24  & 0.37 & -0.41 & 0.41 \\ \hline
G & 32.17 & 32.19 & -0.01 & 0.00 & 0.02  & -0.01 & 0.97 & 0.02  & 0.46 \\ \hline
H & 30.51 & 31.62 & -0.34 & 0.00 & 1.11  & -0.34 & 1.71 & 1.11  & 0.28 \\ \hline
I & 28.57 & 30.73 & -0.58 & 0.00 & 2.16  & -0.58 & 2.46 & 2.16  & 0.13 \\ \hline
\end{tabular}%
}
\label{tab:plate-random-stat}
\end{table}

\begin{table}[h!]
\centering
\caption{Prediction error rate of 2000 samples for models with cutout regions (red color means worse prediction comparing to baseline)}
\resizebox{0.8\textwidth}{!}{%
\begin{tabular}{|c|l|c|c|c|c|}
\hline
\multicolumn{2}{|c|}{\multirow{2}{*}{Prediction Error comparing to Baseline}} &
  \multicolumn{2}{c|}{PC-VR composite} &
  \multicolumn{2}{c|}{HPR-VR composite} \\ \cline{3-6} 
\multicolumn{2}{|c|}{}                      & DiNN-N & DiNN-NC & DiNN-N & DiNN-NC \\ \hline
\multicolumn{2}{|c|}{Ring MER reduction}    & \textbf{---}    & \textbf{---}     & \textcolor{red}{-21\%}  & 33\%    \\ \hline
\multicolumn{2}{|c|}{Matrix MER reduction}  & 60\%   & 79\%    & 3\%    & 59\%    \\ \hline
\multicolumn{2}{|c|}{Contour MSE reduction} & \textcolor{red}{-133\%} & 86\%    & \textcolor{red}{-84\%}  & 88\%    \\ \hline
\end{tabular}%
}
\label{tab:pred_error_cut_2000}
\end{table}

\begin{table}[h!]
\centering
\caption{Prediction error rate of 2000 samples for models without cutout regions}
\resizebox{0.8\textwidth}{!}{%
\begin{tabular}{|c|l|c|c|}
\hline
\multicolumn{2}{|c|}{\multirow{2}{*}{Prediction Error comparing to Baseline}} & SP-VR composite & HP-VR composite \\ \cline{3-4} 
\multicolumn{2}{|c|}{}                      & DiNN-N & DiNN-N \\ \hline
\multicolumn{2}{|c|}{Fiber MER reduction}   & 41\%   & 26\%   \\ \hline
\multicolumn{2}{|c|}{Matrix MER reduction}  & 51\%   & 33\%   \\ \hline
\multicolumn{2}{|c|}{Contour MSE reduction} & 27\%   & 27\%   \\ \hline
\end{tabular}%
}
\label{tab:pred_error_no_cut_2000}
\end{table}

\begin{table}[h!]
\centering
\caption{Prediction error rate of 2000 samples for models with spatial randomness (red color means worse preiction comparing to baseline)}
\resizebox{0.8\textwidth}{!}{%
\begin{tabular}{|c|c|c|c|c|}
\hline
\multicolumn{2}{|c|}{\multirow{2}{*}{Prediction Error comparing to Baseline}}                            & \multicolumn{2}{c|}{PC-SR composite} & SP-SR composite \\ \cline{3-5} 
\multicolumn{2}{|c|}{}   & DiNN-N   & DiNN-NC  & DiNN-N  \\ \hline
\multirow{3}{*}{\begin{tabular}[c]{@{}c@{}}Controlled\\ randomness\end{tabular}}   & Fiber MER reduction & \textbf{---}               & \textbf{---}              & 23.81\%         \\ \cline{2-5} 
 & Matrix MER reduction  & \textcolor{red}{-68.31\%} & 54.81\%  & 45.58\% \\ \cline{2-5} 
 & Contour MSE reduction & \textcolor{red}{-4.26\%}  & \textcolor{red}{-3.73\%}  & 32.26\% \\ \hline
\multirow{3}{*}{\begin{tabular}[c]{@{}c@{}}Uncontrolled\\ randomness\end{tabular}} & Fiber MER reduction & \textbf{---}               & \textbf{---}              & 16.08\%         \\ \cline{2-5} 
 & Matrix MER reduction  & \textcolor{red}{-48.11\%} & \textcolor{red}{-18.37\%} & \textcolor{red}{-3.88\%} \\ \cline{2-5} 
 & Contour MSE reduction & \textcolor{red}{-7.99\%}  & \textcolor{red}{-23.19\%} & 21.05\% \\ \hline
\end{tabular}%
}
\end{table}

\end{document}